\documentclass[11pt]{article}%
\usepackage{amsfonts}
\usepackage{amssymb}
\usepackage{graphicx}
\usepackage{setspace}
\usepackage[round]{natbib}
\usepackage{amsmath}%
\usepackage{amsthm}%
\usepackage{graphicx}
\usepackage{bbm}
\usepackage{palatino}
\usepackage{paralist}
\usepackage{multirow}
\usepackage{booktabs}
\usepackage{dsfont}
\usepackage{indentfirst}
\usepackage{enumerate}
\usepackage{enumitem}
\usepackage{longtable}
\usepackage{bm}
\usepackage{mwe}
\usepackage{float}
\usepackage{booktabs}
\usepackage{lscape}
\usepackage[hyperindex,breaklinks]{hyperref}
\hypersetup{colorlinks=true,       % false: boxed links; true: colored links
	linkcolor=red,       
	citecolor=blue,        % color of links to bibliography
	filecolor=magenta,      % color of file links
	urlcolor=cyan           % color of external links
}  
\usepackage{multirow}
\newtheorem{theorem}{Theorem}
\newtheorem{remark}{Remark}
\theoremstyle{definition}
\newtheorem{definition}{Definition}[section]

\newtheorem{proposition}{Proposition}

\title{Data Augementation with P\'olya Inverse Gamma}

\author{
	\makebox[.4\linewidth]{Jingyu He}\\
	\textit{\small  Department of Management Sciences}\\
	\textit{\small  City University of Hong Kong}
	\and
	\makebox[.4\linewidth]{Nicholas Polson}\\
	\textit{\small  Booth School of Business}\\
	\textit{\small  The University of Chicago}\\
	\and 
	\makebox[.4\linewidth]{Jianeng Xu}\\
	\textit{\small  Booth School of Business}\\
	\textit{\small  The University of Chicago}\\
}

\begin{document}
	\maketitle

%%%%%%%%%%%%%%%%%%%%%
\begin{abstract}
%%%%%%%%%%%%%%%%%%%%%

\noindent We use the theory of normal variance-mean mixtures to derive a data augmentation scheme for models that include gamma functions. Our methodology applies to many situations in statistics and machine learning, including  Multinomial-Dirichlet distributions, Negative binomial regression, Poisson-Gamma hierarchical models, Extreme value models, to name but a few.  All of those models  include  a gamma function  which does not admit a natural conjugate prior distribution providing a significant challenge to inference and prediction.   To provide a data augmentation strategy, we construct and develop the theory of  the class of P\'olya Inverse Gamma distributions. This allows scalable EM and MCMC algorithms to be developed. We  illustrate our methodology on a number of examples, including  gamma shape inference, negative binomial regression and  Dirichlet allocation. Finally, we conclude with directions for future research. 

\end{abstract}

\newpage
\section{Introduction}
\label{sec:Introduction}
Statistical models involving gamma functions are prevalent in statistics and machine learning. For example, topic models, negative binomial regression, time series count models, Poisson-Gamma hierarchical models, non-parametric Bayes,  to name but a few \citep*{rossell2009gaga,aktekin2018sequential,lijoi2016innovation}. Gamma distribution also serves as the conjugate prior for many model parameters, such as a normal precision and Poisson intensity. The normalizing constant depends on gamma function whose argument is the shape parameter. Bayesian Inference in gamma models is a long standing problem that presents significant technical and computational difficulties \citep*{damsleth1975conjugate, rossell2009gaga, miller2018fast}. Similar issue also occurs to the learning of other widely-used gamma models. Table \ref{tab:gamma model} gives a list of distributions, where the shape/concentration/dispersion parameters are nested in gamma functions.

By exploiting normal variance-mean mixture  identities related to gamma function, we derive a general data augmentation strategy. Our main result is given in Proposition \ref{prop:joint}, according to which a MCMC algorithm is built in Proposition \ref{prop:cond} and an expectation-maximization algorithm is developed in Section \ref{sec:em}.

The main difficulty is to address the reciprocal gamma function $1/\Gamma(\alpha)$. Following \cite*{barndorff1982normal}, \cite*{hartman1976completely} and \cite*{roynette2005couples}, we represent $1/\Gamma(\alpha)$ as a mean-variance mixture of normals, then we define our new class of distributions named the P\'olya Inverse Gamma (P-IG)\footnote{It was previously circulated under the name of Exponential Reciprocal Gamma} class. This novel distribution places gamma models in the same footing as other commonly used Bayesian models, such as sparsity (lasso, horseshoe) and logit (P\'olya-Gamma). As a by-product, we show how to nest the latter within our framework. Thus we unify the inference procedure for many models.

Our data augmentation strategy with P-IG auxiliary variables may be utilized to design efficient Markov chain Monte Carlo (MCMC) algorithms in latent Dirichlet allocation \citep*{blei2003latent}, Beta-negative binomial models \citep*{zhou2012beta}, and Gamma-Gamma (GaGa) hierarchical models \citep*{rossell2009gaga}.  It adds to the literature on Bayesian computation with auxiliary variables, which have proven useful in computing posterior distributions in logistic regression \citep*{polson2013bayesian}, negative binomial regression \citep*{pillow2012fully}, multinomial factor models \citep*{holmes2006}, support vector machines \citep*{mallick2005bayesian, polson2011data}, and dependent multinomial models \citep*{linderman2015dependent}.  

To illustrate our methodology, we show examples including gamma shape inference, negative binomial regression and multinomial-Dirichlet model. The first example has a posterior density which exactly matches with the above form. Hence our algorithms can be straightforwardly applied. The posterior samples are efficiently drawn and the posterior mode is easily found by EM algorithm. The second example extends our results to incorporate P\'olya-Gamma mixture representation described in \citep*{polson2013bayesian}. The last example demonstrates how our methodology generalizes to high-dimensional case. Conditioned on the auxiliary variables, all elements of the multivariate concentration parameter become mutually independent, which  reduces the sampling difficulty significantly. We also suggest the use of normal approximation to speed up the sampling procedure in this example.

\begin{table}[]
	\caption{List of Gamma Models}
	\small
	\centering
	\label{tab:gamma model}
	\begin{tabular}{@{}lcc@{}}
		\toprule
		Distribution  & Likelihood  &  Applications \\ \midrule
		Gamma         & $\frac{\beta^\alpha}{\Gamma(\alpha)} x^{\alpha-1}e^{-\beta x}$ &Gamma process, Poisson regression \\  
		& &  \\
		Inverse Gamma & $\frac{\beta^\alpha}{\Gamma(\alpha)} x^{-\alpha-1}e^{-\beta/x}$  & survival analysis, conjugate prior\\ 
		& &  \\
		Beta & $\frac{\Gamma(\alpha+\beta)}{\Gamma(\alpha)\Gamma(\beta)} x^{\alpha-1}(1-x)^{\beta-1}$& order statistics, wavelet analysis \\ 
		& &  \\
		Dirichlet & $\frac{\Gamma(\sum_{k=1}^{K}\alpha_k)}{\prod_{k=1}^{K}\Gamma(\alpha_k)}\prod_{k=1}^K x_k^{\alpha_k-1}$& topic model, Bayesian networks\\
		& &  \\
		Negative Binomial  & $\frac{\Gamma(x+\alpha)}{\Gamma(\alpha)\Gamma(\alpha+1)}p^r(1-p)^{x}$& stock control problems,  negative binomial regression\\
		\bottomrule
	\end{tabular}
\end{table}

The P-IG family of distributions is defined as an infinite convolution of Generalized inverse Gaussian (GIG) distributions and is related to the class of P\'olya-Gamma (PG) distributions \citep*{polson2013bayesian} for logistic regression. Other mixture representations related to GIG are introduced in \cite*{zhang2012ep} and \cite*{barndorff2012basics}, with applications in sparse regression and stochastic volatility modelling. The P-IG(0) distribution is also a special case of $H_a^\Gamma$ distribution family studied in \cite*{roynette2005couples}, who provide a representation of the ratio gamma functions as a scale mixture of normals. This adds to scale mixtures results in Bayesian inference, see \cite*{andrews1974scale}, \cite*{barndorff1982normal}, \cite*{west1987scale}, and \cite*{polson2013bayesian}. Scale mixtures of normals are increasingly used in modeling complex high-dimensional distributions, and \cite*{bhattacharya2016fast} provide fast sampling strategies, adding to the practical use of scale mixture distributions in scalable stochastic simulations.  Equivalently constructed scalable PG sampling schemes are provided in \cite*{windle2014sampling} and \cite*{glynn2018bayesian}. 

\subsection{Connections with Existing Work}
Obtaining random draws or finding the mode from a posterior distribution involving gamma functions is computationally challenging it requires accurate gamma function evaluations. Approximation methods have been proposed to handle the computational burden. \cite*{minka2000estimating} describes an efficient iterative schemes for maximum likelihood estimate of Dirichlet distribution. The likelihood of Dirichlet precision is approximated by a simpler function (gamma density) by matching the first two derivatives, while the multivariate Dirichlet mean is estimated separately by fixed-point iteration. \cite*{miller2018fast} applies the same idea of derivative-matching and approximate the full conditional distribution of gamma shape parameter by a gamma density function. \cite*{rossell2009gaga} defines a gamma shape distribution in differential expression analysis. By approximating gamma function with Stirling's formula and evaluating the limit of the expression, the proposed distribution is roughly proportional to a gamma density. However, These ad-hoc methods are all essentially derived in univariate case. It's not straightforward to generalize them in multivariate cases as we need to deal with correlations, and the computation of approximating parameters itself will get cumbersome. Our framework instead provides an easy and unified way to derive both MCMC and EM algorithms for multivariate models, without the need to approximate density functions. 

The rest of our paper proceeds as follows: Section \ref{sec:P-IG} defines the class of P-IG distributions; Section \ref{sec:DA} illustrates our data augmentation strategy, developing a parameter expanded Gibbs sampler as well as EM algorithm; Section \ref{sec:examples} presents examples of gamma shape inference, negative binomial regression and multinomial-Dirichlet model; and Section \ref{sec:discussion} concludes with directions for future research. 

\section{P\'olya Inverse Gamma (P-IG) Distribution}\label{sec:P-IG}
In this section, we present the theoretical development of the P-IG distribution class, defining the P-IG distribution by the form of its integral transform. In Section \ref{sec:P-IG(0)}, we define a baseline case of the P-IG distribution and prove that it is an infinite convolution of independent GIG distributions;
in Section \ref{sec:P-IG(c)}, the general class of exponential reciprocal distributions is constructed with an exponential tilting of the baseline case defined in Section \ref{sec:P-IG(0)}. Other convolutions of GIG distribution are discussed in Section \ref{sec:GIG}.
\subsection{P-IG(0) Distribution}\label{sec:P-IG(0)}
Let P-IG$(0)$ denote the P\'olya Inverse Gamma distribution. The parameter, which is a tilting parameter fixed at zero in this case, will be discussed in greater detail in Section \ref{sec:P-IG(c)}.
\begin{definition}
	Random variable $X_0$ has an P\'olya Inverse Gamma distribution, P-IG$(0)$, if and only if its density function $p_0(x)$ satisfies the following identity
	\begin{equation}\label{LT_b_0}
	E\left( e^{-s^2 X_0}\right) = \int_{0}^\infty e^{- s^2x} p_0(x) d x = \frac{e^{-\gamma s}}{\Gamma(1+s)} , \quad  \quad  s > 0 .
	\end{equation}
	where $\gamma = -\psi(1) \approx 0.57721$ is the Euler-Mascheroni constant and $\psi(s) = \frac{d}{ds}\log\Gamma(s)$ is the digamma function. 
\end{definition}

\begin{remark}
	\label{rmk:remark_1}
	The product representation for the reciprocal gamma function due to Weierstrass is,
	$$
	\frac{e^{-\gamma s}}{\Gamma(1+s)} = \prod_{k=1}^\infty \left(1 + \frac{s}{k}\right) e^{-\frac{s}{k}}, \, s \in \mathbb{C}\slash \{0, -1, -2, ...\}
	$$
\end{remark}

\begin{remark}
	\cite*{roynette2005couples} prove the existence of an infinitely divisible distribution $H_a^\Gamma $, with density function  $  p_a^\Gamma  ( x )$, such that for $a>0$
	$$
	E\left( e^{-\frac{1}{2}s^2 H_a^\Gamma}\right) = \int_0^\infty e^{-\frac{1}{2} s^2 x} p_a^\Gamma  ( x ) dx =\frac{\Gamma(a)}{\Gamma(a+s)} e^{\psi(a)s} 
	$$
	which follows from the general product representation for the reciprocal gamma function, 
	\begin{equation}
	\frac{\Gamma(a)}{\Gamma\left(a + s\right)} e^{\psi(a)s} =\prod_{k=0}^{\infty}\left( 1 + \frac{s}{a+ k} \right)e^{-\frac{s}{a+k}}.
	\end{equation}
	Note that the representation in Remark \ref{rmk:remark_1} is a special case with $a = 1$. Hence $$\text{P-IG}(0) \overset{D}{=} \frac{1}{2} H_1^\Gamma.$$
\end{remark}

\begin{remark}
	The P-IG($n,0$) is defines as follows. If $X_{n,0} \sim \text{P-IG}(n,0)$,
	\begin{equation*}
	E\left( e^{-s^2 X_{n,0}}\right) = \int_{0}^\infty e^{- s^2x} p_{n,0}(x) d x = \frac{e^{-\gamma n s}}{\Gamma(1+s)^n} , \quad  \quad n>0, s > 0 .
	\end{equation*}
	Note that $\text{P-IG}(n,0)$ is the equivalent in distribution to the sum of $n$ independent $\text{P-IG}(1,0)$ when $n$ is a positive integer.
\end{remark}

\subsection{General P-IG(c) Distribution}\label{sec:P-IG(c)}
\label{subsec:PIG_dc}
We now construct the general class of P-IG distributions, P-IG$(c)$, by exponentially tilting the P-IG$(0)$ distribution.  The exponential tilting strategy -- similar to the one used by \cite*{polson2013bayesian} -- allows a second parameter $c \in \mathbb{R}^+$ to inform a priori the precision of the P-IG random variable.        
\begin{definition}
	\label{def:PG_dc}
	The P-IG$(c)$ distribution is constructed as an exponential tilting of the P-IG$(0)$ density. Its density function is
	\begin{equation*}
	p_c\left(x\right) = Z_c \cdot \exp\left(- c^2 x\right) p_0(x),\quad   x, c >0.
	\end{equation*}
	The normalizing constant, namely $Z_c = 1/E\left(\exp\left(-c^2 X_0\right)\right)$ where $X_0$ is an P-IG$(0)$ random variable, can be calculated using the identity in Remark \ref{rmk:remark_1}. Similarly, the integral identity of P-IG$(c)$ is given by
	\begin{align}
	E\left(e^{-s^2 X_c}\right) &= \prod_{k=1}^\infty \left( \frac{k + \sqrt{s^2 + c^2}}{k + c} \right) e^{-\frac{\sqrt{s^2 + c^2}-c}{k}}\\
	&= \frac{\Gamma(1+c)}{\Gamma(1+\sqrt{s^2+c^2})} e^{-\gamma(\sqrt{s^2+c^2}-c)}. \label{def:general}
	\end{align}
\end{definition}

Our main result, presented in Theorem \ref{thm:theorem_1}, is that a random variable $X_c\sim$ P-IG$(c)$ may be constructed from an infinite sum of independent generalized inverse Gaussian (GIG) random variables.  The power of the result lies in the ability to identify previously unknown conditional posterior distributions.

\begin{theorem}
	\label{thm:theorem_1}
	The P-IG$(c)$ class of distributions can be constructed as an infinite sum of independent generalized inverse Gaussian (GIG) distributions as follows
	$$
	\text{P-IG}(c) \overset{D}{=} \sum_{k=1}^\infty \text{GIG}\left(-\frac{3}{2}, 2c^2, \frac{1}{2k^2}\right).
	$$ 
	In particular, when $c=0$, the GIG distribution reduces to inverse gamma distribution. Hence,
	\begin{align*}
	\text{P-IG}(0) \overset{D}{=} \sum_{k=1}^\infty \frac{1}{4k^2} \Gamma^{-1}_k.
	\end{align*}
	where $\Gamma^{-1}_k$ are i.i.d. inverse gamma random variables with shape $\frac{3}{2}$ and scale 1. 
\end{theorem}
\begin{proof}
	See Appendix \ref{appendix:proof}.
\end{proof}
The following theorem concerns the first two moments of P-IG($c$) which will be used to construct our EM algorithm in Section \ref{sec:em} and the approximate Gibbs sampler in Appendix \ref{appendix:aGibbs}.
\begin{theorem}
	\label{thm:moments}
	If $G_k$ is a GIG random variable with $p = -\frac{3}{2}, a=2c^2, b=\frac{1}{2k^2}$, then the mean and variance of the tail infinite sum $\sum_{k=N}^\infty G_k$ are
	\begin{align*}
	E\left(\sum_{k=N}^\infty G_k\right) &= \frac{1}{2c}\left(\psi(N+c) - \psi(N)\right) \\
	\text{Var}\left(\sum_{k=N}^\infty G_k\right) &= \frac{1}{4c^3} \left(\psi(N+c) - \psi(N) - c \psi'(N+c)\right), 
	\end{align*}
where $\psi(s) = \frac{d}{ds}\log\Gamma(s)$ is the digamma function. Setting $N=1$ gives us the first two moments of P-IG($c$).
\end{theorem}
	\begin{proof}
	If $g_k$ is a GIG random variable with $p = -\frac{3}{2}, a=2c^2, b=\frac{1}{2k^2}$, then its mean and variance are given by
	\begin{equation*}
	E(g_k) = \frac{1}{2}\left(\frac{1}{k^2+ck}\right), \quad 
	\text{Var}(g_k) = \frac{1}{4c} \left(\frac{1}{k^3+2ck^2+c^2k}\right).
	\end{equation*}
	Theorem \ref{thm:moments} is thus a direct application of Theorem \ref{thm:theorem_1}.
\end{proof}
\begin{remark}
	The P-IG$(c)$ distribution class belongs to the family of generalized gamma convolutions (GGC), \cite*{bondesson1992generalized}. Its Laplace transform in Equation (\ref{def:general}) also satisfies
	\begin{equation*}
	E(e^{-sX_c}) = \exp\left\{\int_0^\infty (e^{-sx}-1) \nu(x)dx \right\},\quad  s\geq 0.
	\end{equation*}
	Its L\'evy density $\nu(x)$ and Thorin density $\mu(t)$ are 
	\begin{align*}
	\nu(x) &= \frac{1}{x} \int_{0}^\infty e^{-tx} \mu(t) dt,\\
	\mu(t) &= \mathds{1}_{t\geq c^2}\frac{\psi(1-\sqrt{c^2-t}) + \psi(1+\sqrt{c^2-t})+ 2\gamma}{4\pi \sqrt{t-c^2}}.
	\end{align*}
	This result can be used to generate P-IG random variables as it shows that it falls into the class of Generalized Gamma Convolution, see \cite*{bondesson1982simulation} and \cite*{rosinski2001series}.
\end{remark}

\subsection{GIG Mixtures}\label{sec:GIG}
The P-IG distribution allows us to represent the unnormalized density $\frac{e^{a x}}{\Gamma(1+x)}$ as a normal variance-mean mixture. That is, 
\begin{align*}
\frac{e^{a x}}{\Gamma(1+x)} = \int_0^\infty \phi(x \mid \mu(\omega), \sigma^2(\omega)) \cdot \tau(\omega)p_0(\omega) d\omega 
\end{align*}
where $\phi(\cdot \mid \mu, \sigma^2)$ is the normal density with mean $\mu$ and variance $\sigma^2$ and $P_0(\cdot)$ is the distribution function of P-IG(0). $\mu(\omega) = (a+\gamma)/(2\omega), \sigma^2(\omega) = 1/(2\omega)$ and $\tau(\omega) = \sqrt{\pi/\omega}\exp\left\{(a+\gamma)/4\omega\right\}$.  In statistics and machine learning, probability distributions whose density function $p(x)$ is of the following form are used explicitly and implicitly:
\begin{equation}
p(x) = \int_{0}^\infty f(x \mid \theta(\omega)) p(\omega) d\omega.
\end{equation}
Here $f(x \mid \theta(\omega))$ is some well-known density function, e.g. normal, and the mixing $p(\omega)$ is the distribution of a single GIG or an infinite convolution of GIG's. Combined with a data-augmentation scheme, the above mixture representation provides a powerful framework to solve many non-Gaussian models.
\begin{enumerate}
	\item When $f = \phi$ and the mixing distribution is GIG, \cite*{polson2013data} give the variance-mean mixture representations for many common loss functions in regression and binary classification problems, which corresponds to different choices of the function $\theta(\omega) = (\mu(\omega), \sigma^2(\omega))$ and parameters of GIG. For example, absolute loss $L(y)=|y|$, hinge loss for support vector machine $L(y) = \max(1-y, 0)$ and check loss for quantile regression $L(y)=|y|+(2q-1)y$. The representations then help reduce those non-Gaussian models to Gaussian linear models with heteroscedastic errors. Note that GIG is a very general family with many common distributions as its special cases, such as gamma, inverse gamma and inverse Gaussian distribution. 
	\item For the logistic loss in binary classification $L(y) = \log(1+e^y)$, \cite*{polson2013data} show that it is also a normal variance-mean mixture. The mixing distribution is P\'olya distribution, which is constructed as an infinite sum of exponentials. Note that exponential distribution is again a special case of GIG.
	\item By choosing $f$ to be the exponential power density, $f(x \mid \eta(\omega), q) \propto \exp\left\{-\frac{1}{2\eta(\omega)}|x|^q\right\}$, and the mixing distribution $p(\omega)$ to be GIG,  \cite*{zhang2012ep} introduce a sparsity-inducing prior called EP-GIG and develop EM algorithms for sparse learning. The density function of EP-GIG is given explicitly and special cases (generalized $t$ distribution and exponential power-gamma distribution) are discussed when the mixing GIG reduces to inverse gamma and gamma respectively.
	\item The P\'olya-Gamma distribution class is proposed by \cite*{polson2013bayesian} to solve the inference problem in models with binomial likelihoods, including logistic regression and negative binomial regression.  P\'olya-Gamma distribution PG$(b,c)$ can be written as an infinite convolution of gamma distributions whose shapes are all equal to $b$ and scales depend on $c$, or equivalently $\text{GIG}(b,2,0)$. 
	\begin{equation*}
	\text{PG}(b,c) \overset{D}{=} \frac{1}{2\pi^2} \sum_{k=1}^\infty \frac{\text{GIG}(b,2,0)}{(k-1/2)^2 + c^2/(4\pi^2)}.
	\end{equation*} 
	Choose $f(z \mid \omega) = 2^{-b} \exp\left\{-\omega z^2/2 + \kappa z\right\}$ with $\kappa = a-b/2$ and $z = x'\beta$, the likelihood of logistic regression is a mixture 
	\begin{equation}\label{eqn:pg integral}
	\frac{(e^z)^a}{(1+e^z)^b} = \int_0^\infty f(z \mid \omega)p_{\text{PG}}(\omega \mid b,0) d\omega.
	\end{equation}	
	Here $p_{\text{PG}}(\omega \mid b,0)$ is the density of PG$(b,0)$ and $f(z \mid \omega)$ is proportional to the normal density where the mean and variance are functions of $\omega$.
	\item \cite*{barndorff2012basics} use a normal variance-mean mixture as a general approach of building densities on the real line. Here $f(x \mid \theta(\omega)) = \phi(x \mid \mu+\beta\omega, \omega)$. When $p(\omega)$ is GIG density, the resulted mixture is generalized hyperbolic distribution, which includes many special cases such as normal inverse Gaussian, normal gamma, Laplace, skewed Student's $t$ distribution. Furthermore, normal distribution can also be written as a limiting case of generalized hyperbolic distribution. 
\end{enumerate}

\section{MCMC and Data Augmentation}\label{sec:DA}
This section illustrates the data augmentation strategy and sampling scheme for Gamma inference using the P-IG class distributions. First, notice that many Bayesian gamma models involve a posterior density of the form
\begin{equation}\label{eqn:post}
p(x)  = C \cdot \left(\prod_{\ell=1}^L \Gamma(g_\ell(x)) \right) \left(\prod_{m=1}^M \frac{1}{\Gamma(h_m(x))}\right) \left(\prod_{n=1}^N\frac{\Gamma(j_n(x))}{\Gamma(j_n(x) + \beta_n)}\right) \cdot x^{p-1} e^{-ax^2 + bx}, \; x>0,
\end{equation}  
where $x \in \mathbb{R}^+$ is on the positive real line, and $C$ is the normalizing constant. The arguments in gamma functions, $\left\{g_\ell(\cdot)\right\}_{\ell=1}^L, \left\{h_m(\cdot)\right\}_{m=1}^M$ and $\left\{j_n(\cdot)\right\}_{n=1}^N$ are nonnegative increasing linear functions of $x$ on $(0,\infty)$. We assume $M \geq 1$, otherwise the form might not be integrable. Parameters $p, a, b, \beta_1, ..., \beta_N$ are scalars. $\beta_n$'s are positive. $p$ and $a$ are nonnegative.  In order to perform full posterior inference on the variable $x$, a sampling procedure for $p(x)$ is needed, while a Maximum A Posteriori (MAP) estimate requires finding the maximizer of it. 

The idea of the data augmentation strategy is to introduce a group of auxiliary random variables $\bm{\omega} = (\omega_1, ..., \omega_n)^T$ such that for $\bm{\omega} \in \Omega$,
$$
p(x) = \int_\Omega p(x,\bm{\omega}) d\bm{\omega}
$$ 
and the joint density $p(x,\bm{\omega})$ after augmentation is easier to deal with, as it doesn't consist of gamma functions any longer.

\begin{table}[]
	\caption{Integral Representation for Gamma Functions}
	\centering
	\label{tab:gamma}
	\begin{tabular}{@{}lcc@{}}
		\toprule
		  & Integral Representation  &  Auxiliary Variables\\ \midrule
		$\Gamma(\alpha)$ & $\int_{0}^{\infty} x^{\alpha-1} e^{-x} dx$& Gamma  \\
		& & \\
		$\frac{1}{\Gamma(\alpha)}$& $\int_{0}^\infty \alpha e^{-\alpha^2x+\gamma\alpha} p_0(x) dx$& P-IG\\
		& &\\
		$\frac{\Gamma(\alpha)}{\Gamma(\alpha+\beta)}$ & $\int_{0}^1 \frac{1}{\Gamma(\beta)} x^{\alpha-1}(1-x)^{\beta-1}dx$& Beta\\
		\bottomrule
	\end{tabular}
\end{table}

Returning to Equation (\ref{eqn:post}), the posterior density involves gamma functions $\Gamma(\cdot)$, reciprocal gamma functions $1/\Gamma(\cdot)$ and gamma ratios $\Gamma(\cdot)/\Gamma(\cdot + \beta)$. We will then express each of them using the corresponding integral representation in Table \ref{tab:gamma}. This is equivalent to introducing an auxiliary random variable for each of them. The total number of the auxiliary random variables is then $L+M+N$.

\begin{proposition}\label{prop:joint}
	The posterior density admits an integral representation as follows
	\begin{equation}\label{eqn:aug-post}
	p(x)  = C\cdot \int_{(0,\infty)^{L+M} \times (0,1)^N} G(x, \bm{\tau})\cdot H(x, \bm{\omega}) \cdot J(x, \bm{\eta}) \cdot Q(x) \cdot e^{-ax^2 + bx} d\bm{\tau}d\bm{\omega}d\bm{\eta} 
	\end{equation}
	where 
	\begin{align*}
	G(x, \bm{\tau}) &= \prod_{\ell=1}^L \tau_{\ell}^{g_{\ell}(x)-1} e^{-\tau_\ell} \\
	H(x, \bm{\omega}) &= \exp\left\{-\sum_{m=1}^M h_m(x)^2 \omega_m + \gamma\sum_{m=1}^M h_m(x)\right\} \prod_{m=1}^M p_0(\omega_m) \\
	J(x, \bm{\eta}) &= \prod_{n=1}^N \eta_n^{j_n(x)-1} (1-\eta_n)^{\beta_n-1}\\
	Q(x) &= x^{p-1} \prod_{m=1}^M h_m(x).
	\end{align*}
	$p_0(\cdot)$ is the probability density of P-IG(0).
\end{proposition}

\begin{remark}
	When $N\geq 1$ in the form (\ref{eqn:post}), the third term $\left(\prod_{n=1}^N\frac{\Gamma(j_n(x))}{\Gamma(j_n(x) + \beta_n)}\right)$ can be absorbed into the first two terms. However, we still recommend using the Beta representation if possible. Otherwise the total number of auxiliary variables needed is increased by $N$.
\end{remark} 

\begin{remark}
	We may generalize the form (\ref{eqn:post}) in a few ways:
	\begin{enumerate}
		\item For multivariate $x$ of dimension $d$, if $g,h,j$ are all linear functions mapping from $\mathbb{R}^{+, d}$ to $\mathbb{R}^+$, then the data augmentation is still valid.
		\item The $x^p$ term can be replaced with a polynomial function of $x$, as long as it's always positive for $x > 0$.
		\item If the posterior density $p(x)$ has extra factors which are not included in the form (\ref{eqn:post}), the strategy still works as long as we can find integral representations for those extra factors.
	\end{enumerate} 
\end{remark}

Although the posterior joint distribution in Equation (\ref{eqn:aug-post}) looks forbidding at the first glance, in many applications the linear functions $g,h,j$ are simple enough, e.g. $h_m(x)=x$ for all $m$, and $N=0$, which simplifies the expression a lot. More importantly, the conditional posteriors can be derived easily. To derive  that of $\tau_\ell$, for example,
$$
p(\tau_\ell \mid x, \bm{\tau}_{-\ell}, \bm{\omega,\eta}) = \tau_{\ell}^{g_{\ell}(x)-1} e^{-\tau_\ell} \cdot  \frac{\left(\prod_{\ell'\neq\ell} \tau_{\ell'}^{g_{\ell'}(x)-1} e^{-\tau_\ell'} \cdot H(x, \bm{\omega}) \cdot J(x, \bm{\eta}) \cdot Q(x) \right)}{\int G(x, \bm{\tau})\cdot H(x, \bm{\omega}) \cdot J(x, \bm{\eta}) \cdot Q(x) d\tau_\ell },
$$  
notice that it's proportional to $\tau_{\ell}^{g_{\ell}(x)-1} e^{-\tau_\ell}$. Therefore the conditional posterior distribution of $\tau_{\ell}$ is $\Gamma(g_\ell(x),1)$. Similarly for $\omega_m$'s and $\eta_n$'s. For the conditional posterior of $x$, since $g,h,j$ are all linear, it's proportional to $Q_1(x)e^{-Q_2(x)}$ where $Q_1(x)$ is a polynomial of degree $p+M-1$ and $Q_2(x)$ is a quadratic function.

Before we summarize the conditional posteriors with respect to the augmented vector $(x, \bm{\tau,\omega, \eta})$, the following definition of power truncated normal (PTN) distribution is useful.
\begin{definition}
	The power truncated normal distribution, PTN$(p,a,b)$, has density function
	\begin{equation}
	p(x) = C\cdot x^{p-1} e^{-ax^2+bx}, x>0
	\end{equation}
	where $p,a > 0$ and $b \neq 0$.
\end{definition}
Finally, the following proposition is helpful in developing the Gibbs sampler.
\begin{proposition}[Gibbs Sampler]\label{prop:cond}
	If the joint probability density of the augmented vector $(x, \bm{\tau,\omega, \eta})$ is proportional to $G(x, \bm{\tau})\cdot H(x, \bm{\omega}) \cdot J(x, \bm{\eta}) \cdot Q(x)$ given in Proposition \ref{prop:joint}, then the conditional distributions are 
	\begin{align*}
	\tau_{\ell} \mid x, \bm{\tau}_{-\ell}, \bm{\omega,\eta} &\sim \Gamma(g_\ell(x),1), \quad  \ell = 1, 2, ..., L \\
	\omega_m \mid x, \bm{\tau}, \bm{\omega}_{-m}, \bm{\eta} &\sim \text{P-IG}(h_m(x)), \quad  m = 1, 2, ..., M\\
	\eta_n \mid x, \bm{\tau}, \bm{\omega}, \bm{\eta}_{-n} &\sim \text{Beta}(j_n(x), \beta_n), \quad  n = 1, 2, ..., N\\
	x \mid \bm{\tau}, \bm{\omega}, \bm{\eta} &\sim \sum_{m=0}^M \pi_k \cdot \text{PTN}(p+m, \tilde{a}, \tilde{b})
	\end{align*}
	Here the conditional distribution of $x$ is a finite discrete mixture of PTN distributions. $\{\pi_m\}_{m=0}^M$ are proportional to the coefficients in the polynomial $Q(x)$.
	\begin{align*}
	\tilde{a} &= a + \sum_{m=1}^M \omega_m\cdot(h_m'(0))^2\\
	\tilde{b} &= b + \left(\sum_{\ell=1}^{L} g_\ell'(0)\cdot\log\tau_{\ell}\right) + \left(\sum_{m=1}^M h_m'(0)(\gamma -2\omega_m h_m(0))\right) + \left(\sum_{n=1}^N j_n'(0)\cdot\log\eta_n\right). 
	\end{align*}
	Furthermore, given $x$, all auxiliary random variables are conditionally independent. The sampling methods for P-IG and PTN are given in Appendix \ref{appendix:P-IG} and \ref{appendix:PTN}.
\end{proposition}

In many statistical applications, the number of auxiliary random variables grows linearly with the sample size and problem dimension, but the forms of $g,h,j$ are relatively simple. Observing that $\tilde{a}$ and $\tilde{b}$ are both the sum of numerous terms, we may use normal variables to approximate them, by matching the moments, so that the sampling procedure for $(\bm{\tau,\omega,\eta})$ can be skipped in Gibbs sampling. Appendix \ref{appendix:aGibbs} gives the approximate Gibbs sampler when $g_\ell(x)=g(x)$ for all $\ell$ (similarly for $h$ and $j$), and $L,M,N$ are all large enough.

\subsection{Expectation-Maximization Algorithm}\label{sec:em}
By exploiting Proposition \ref{prop:joint}, MAP of the model can be found using an expectation-maximization algorithm. The complete-data log posterior is 
\begin{equation}\label{eqn:log_post} 
\begin{aligned}
&\log p(x, \bm{\tau,\omega, \eta}) \\ 
&= c_0 + \log G(x, \bm{\tau})  + \log H(x, \bm{\omega}) + \log J(x, \bm{\eta}) + \log Q(x) -ax^2 + bx \\
%&= c_0 + \left(\sum_{\ell=1}^L (g_{\ell}(x)-1)\log\tau_\ell - \tau_{\ell} \right) + \left(\sum_{m=1}^M \gamma h_m(x) - h_m(x)^2 \omega_m + \log p_0(\omega_m) \right)  \nonumber\\
%&+ \left(\sum_{n=1}^N (j_n(x)-1)\log \eta_n + (\beta_n-1)\log(1-\eta_n)\right) + \log Q(x) \\
&= c_1 + \left(\sum_{\ell=1}^L g_{\ell}(x)\log\tau_\ell \right) + \left(\sum_{m=1}^M \gamma h_m(x) - h_m(x)^2 \omega_m \right)  \\
& \quad\quad + \left(\sum_{n=1}^N j_n(x)\log \eta_n \right) + \log Q(x) -ax^2 + bx
\end{aligned}
\end{equation}
for some constants $c_0, c_1$ (with respect to $x$). In the $t$-th expectation step, we compute the expected value of $\log p(x, \bm{\tau,\omega, \eta})$ under the current conditional posterior $p(\bm{\tau,\omega, \eta} \mid x_{(t)})$, denoted as $C(x \mid x_{(t)})$. Then in the maximization step, $C(x \mid x^{(t)})$ is maximized as a function of $x$.  We now derive the expectation and maximization steps.
\begin{itemize}
	\item \textbf{The Expectation Step}\\
	From equation (\ref{eqn:log_post}), notice that $\log p(x, \bm{\tau,\omega, \eta})$ is linear in terms of $\log\tau_\ell, \omega_m$ and $\log\eta_n$. Therefore, we replace them with their conditional expectations in the expectation step. Applying Proposition \ref{prop:cond} yields
	\begin{align*}
	E\left(\log\tau_\ell \mid x_{(t)}\right) &= \psi(g_\ell(x_{(t)})), \quad  \ell = 1, 2, ..., L \\
	E\left(\omega_m \mid x_{(t)}\right) &= \frac{\psi(1+h_m(x_{(t)})) + \gamma}{2h_m(x_{(t)})}, \quad  m = 1, 2, ..., M \\
	E\left(\log\eta_n \mid x_{(t)}\right) &= \psi(j_n(x_{(t)})) - \psi(j_n(x_{(t)}) + \beta_n) , \quad  n = 1, 2, ..., N.
	\end{align*}
	The derivation above uses  the properties of gamma and beta distribution, as well as Theorem \ref{thm:moments} which calculates the expectation of P-IG($c$) distribution. Finally, one can represent the function $C(x \mid x_{(t)})$ (up to a constant) as 
	\begin{align*}
	C(x \mid x_{(t)})=& \left(\sum_{\ell=1}^L g_{\ell}(x) \psi(g_\ell(x_{(t)})) \right) + \left(\sum_{m=1}^M \gamma h_m(x) - h_m(x)^2 \frac{\psi(1+h_m(x_{(t)})) + \gamma}{2h_m(x_{(t)})} \right) \\
	&+ \left(\sum_{n=1}^N j_n(x)\psi(j_n(x_{(t)})) - \psi(j_n(x_{(t)}) + \beta_n) \right) + \log Q(x) -ax^2 + bx
	\end{align*}
	Given the linearity of $g, h, j$ functions, it can be further simplified as 
	\begin{equation}\label{eqn:e-step}
	C(x \mid x_{(t)}) = \log Q(x) - \kappa_1 x^2 + \kappa_2 x
	\end{equation}
    for some constant $\kappa_1 > 0$ and $\kappa_2 \in \mathbb{R}$, depending on $x_{(t)}$.
	\item \textbf{The Maximization Step}\\
	Since $C(x \mid x_{(t)}) \rightarrow -\infty$ as $x \rightarrow \infty$,  we conclude that $C(x \mid x_{(t)})$ as a function of $x$ has a maximizer on $(0, \infty)$, which can be found numerically. Furthermore, when $h_m(0) = 0$ for all $m$, the unique maximizer has a closed form
	\begin{equation}\label{eqn:m-step}
	x^* := \arg\max_{x>0} C(x \mid x_{(t)}) = \frac{\kappa_2 + \sqrt{\kappa_2^2 + 8\kappa_1(p+M-1)}}{4\kappa_1}.
	\end{equation}
\end{itemize}

\section{Examples}\label{sec:examples}
\subsection{Inference for Gamma Shape}
The gamma distribution, parameterized by shape $\alpha$ and rate $\beta$, is a component of many probability models.  For instance, a gamma prior distribution for the precision parameter in Gaussian linear models is quite common.  In fact, Normal-gamma distributions are workhorse models for shrinkage estimation in regression problems \citep*{griffin2010}. Gamma distribution is also widely used in modelling of extreme values, where it serves as the prior for the shape parameter of Pareto distribution (\cite*{arnold1989bayesian}) and helps to construct a quasi-conjugate prior for generalized Pareto distribution (\cite*{diebolt2005quasi}). While a gamma prior distribution for a parameter is common, it is less common to model hyperparameters of the gamma distribution itself as random variables -- particularly the shape parameter, $\alpha$.  Although posterior inference of the rate parameter is straightforward -- since the gamma distribution itself is a conjugate prior for the rate parameter -- posterior inference of the gamma shape parameter is a long-standing problem \citep*{damsleth1975conjugate, damien1995approximate, rossell2009gaga, miller2018fast} and efficient posterior computation remains an open problem.

\cite*{damsleth1975conjugate} discussed two conjugate priors for $\alpha$, the gamma shape parameter. They are called GamCon distributions of type I and type II. Type I assumes the rate $\beta$ is known, while the type II doesn't. In this section, we focus on the latter one and replicate Damsleth's example, showing how to utilize the data augmentation scheme and build  algorithms for posterior inference.

Let's consider the following hierarchical model
\begin{align*}
x \mid \alpha, \beta &\sim \Gamma(\alpha, \beta),\\
\beta \mid \alpha &\sim \Gamma(\delta \alpha+1, \delta\eta),\\
\alpha &\sim \xi_2(\eta/\mu, \delta).
\end{align*} 
Here $\xi_2$ is GamCon distribution of type II. $\eta>\mu>0, \delta > 0$. The probability density of $\xi_2(\mu, \delta)$ is
\begin{equation*}
p(\alpha \mid \mu, \delta) = C_{\mu,\delta}\cdot \frac{\Gamma(\delta \alpha + 1)}{\Gamma(x)^\delta} (\delta\mu)^{-\delta \alpha},\, \alpha>0, \mu>1, \delta>0. 
\end{equation*}
Suppose observations $\bm{x}=(x_1, x_2, ..., x_n)'$ are independently and identically distributed gamma random variables drawn from the above model. The likelihood is 
\begin{equation*}
f(\bm{x} \mid \alpha, \beta) = \prod_{i=1}^n \frac{\beta^\alpha}{\Gamma(\alpha)} x_i^{\alpha - 1} e^{-\beta x_i} \propto \frac{(x_g\beta)^{n\alpha}}{\Gamma(\alpha)^{n}} e^{-nx_g\beta}
\end{equation*}
where $x_g$ is the geometrical mean, $x_g = \left(\prod_{i=1}^{n} x_i\right)^{1/n}$. 

The marginal posterior distribution of $\alpha$ is then calculated as 
\begin{align*}
p(\alpha \mid \bm{x}) &= \int_0^\infty f(\bm{x} \mid \alpha, \beta)p(\beta \mid \alpha)p(\alpha) d\beta \\
\text{where}\quad \quad p(\beta \mid \alpha) &= \frac{(\delta\eta)^{\delta\alpha + 1}}{\Gamma(\alpha\delta + 1)}\beta^{\alpha\delta} e^{-\delta\eta\beta},\, \text{and}\quad \quad  p(\alpha) = C\cdot \frac{\Gamma(\delta\alpha + 1)}{\Gamma(\alpha)^\delta} (\delta\eta/\mu)^{-\delta\alpha}.
\end{align*} 
By construction, the marginal posterior of $\alpha$ given $\bm{x}$ also follows $\xi_2$, with updated parameters $(\eta', \mu', \delta')$. That is, $\alpha \mid \bm{x} \sim \xi_2(\eta'/\mu', \delta')$ and
\begin{align*}
\delta' &= \delta + n,\\
\eta' &= \frac{\delta}{\delta + n}\eta + \frac{n}{\delta + n} x_a, \\
\mu' &= \mu^{\frac{\delta}{\delta + n}}\cdot x_g^{\frac{n}{\delta + n}}.
\end{align*}
where $x_a$ is the arithmetical mean, $x_a = \frac{1}{n}\sum_{i=1}^{n} x_i$. One can observe that $\eta'$ is a weighted arithmetical mean of $\eta$ and $x_a$ with weights $\delta$ and $n$ respectively and $\mu'$ is a weighted geometrical mean of $\mu$ and $x_g$, also with weights $\delta$ and $n$. Here $x_a$ and $x_g$ are two sufficient statistics. $\delta$ is viewed as the prior sample size while $\eta$ and $\mu$ are the prior means. The problem of gamma shape inference is thus translated to that of $\xi_2$ distribution.

\subsubsection{MCMC and EM for $\xi_2$ when $\delta \in \mathbb{N}^+$}
We first develop the Gibbs sampler for a general GamCon distribution of type II, using data augmentation strategy. Suppose $x \sim \xi_2(x; \mu, \delta)$ and $\delta$ is a positive integer. Then the probability density can be rewritten in the form (\ref{eqn:post})
\begin{equation}
p(x \mid \mu, \delta) = C_{\mu,\delta}\cdot \Gamma(\delta x + 1) \left(\prod_{m=1}^M \frac{1}{\Gamma(x)}\right) e^{-\delta\log(\delta\mu) x}
\end{equation}
Hence $L=1$ with $g_1(x) = \delta x+ 1$, $M=\delta$ with $h_m(x) = x$ and $N=0$. $p=1, a=0$ and $b = -\delta\log(\delta\mu)$.

After introducing the auxiliary variables $\tau$ and $\omega_1, \omega_2, ..., \omega_{\delta}$,  Proposition \ref{prop:cond} then immediately gives the conditional posteriors:
\begin{align*}
\tau \mid x, \bm{\omega} &\sim \Gamma(\delta x +1, 1) \\
\omega_1, \omega_2, ..., \omega_\delta \mid x, \tau &\stackrel{i.i.d.}{\sim}  \text{P-IG}(x)\\
x \mid \tau, \bm{\omega} &\sim \text{PTN}\left(\delta+1, \sum_{i=1}^\delta \omega_i, \delta\left(\gamma - \log(\delta\mu/\tau)\right)\right)
\end{align*}

\subsubsection{Damsleth Examples with Non-Informative Prior}
\begin{figure}[h!]
	\centering
	\caption{Histogram of Posterior Samples}
	\label{fig:gammashapeunknownscale}
	\includegraphics[width=0.9\linewidth]{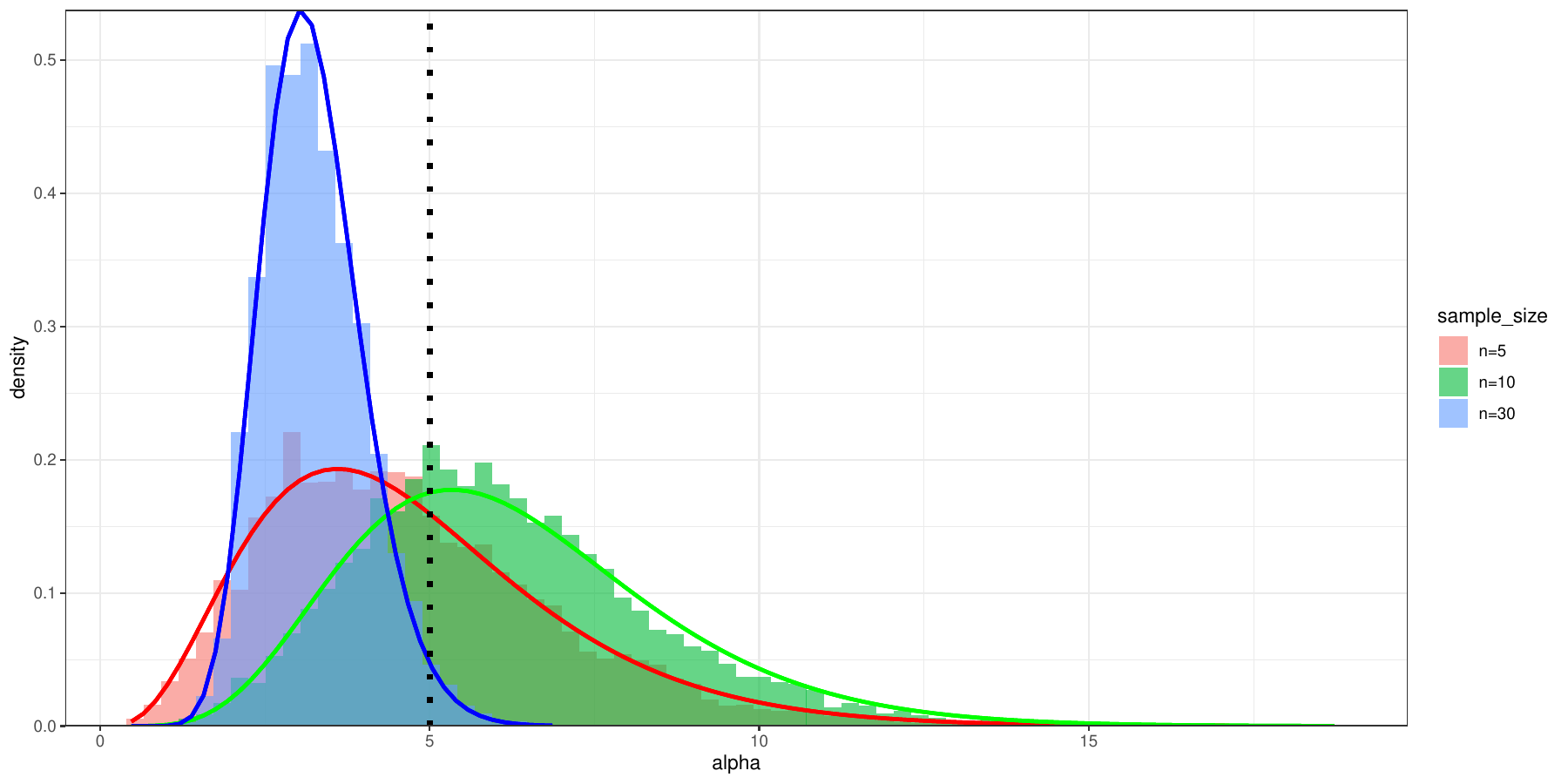}
\end{figure}

\begin{figure}[h!]
	\centering
	\caption{Trace Plot of Posterior Samples}
	\label{fig:gamma_trace}
	\includegraphics[width=0.8\linewidth]{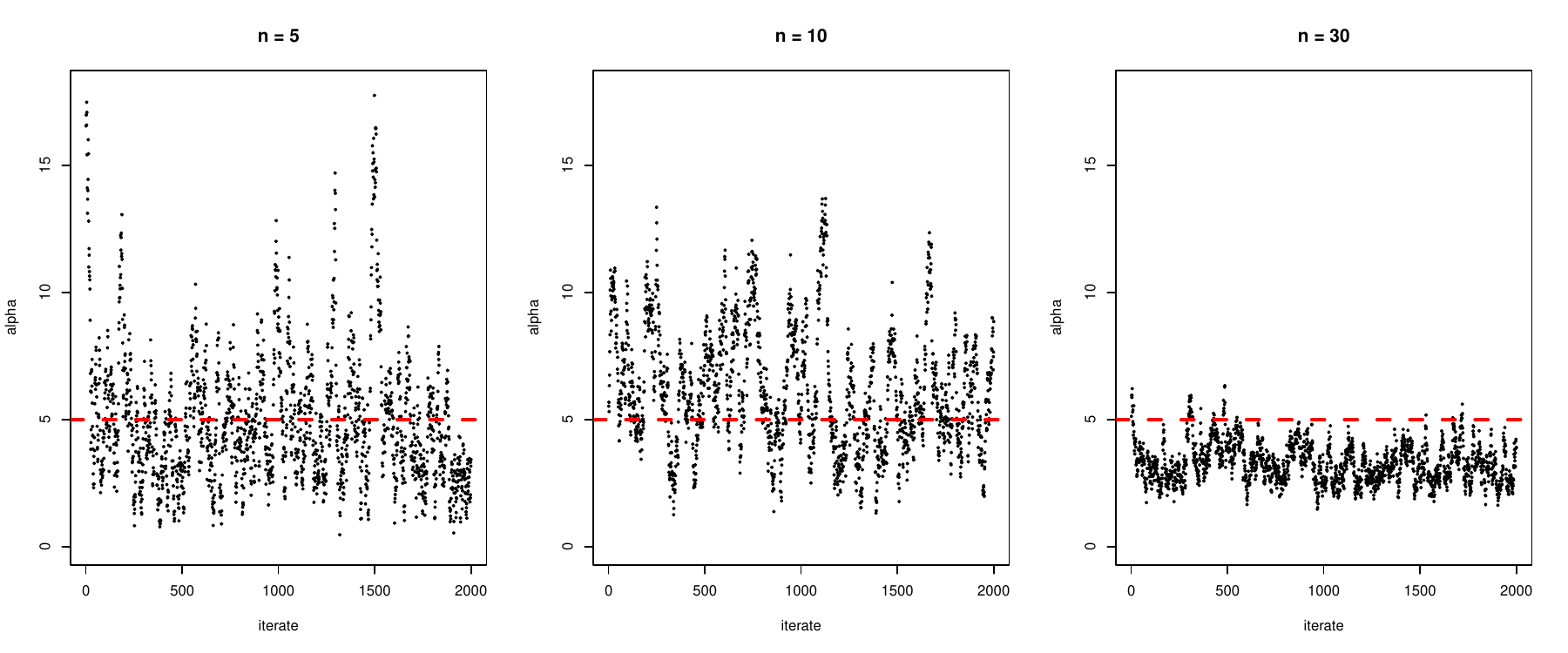}
\end{figure}

Here we replicate Damsleth's example of no prior information with sample size $n=5,10,30$. Putting $\delta = 0$, the posterior parameters are 
\begin{align*}
\delta' = n,\quad  \eta' = x_a,\quad  \mu' = x_g.
\end{align*}
The resulted posterior of gamma shape is thus $\alpha \mid \bm{x} \sim \xi_2(x_a/x_g, n)$. Instead of actually generating Gamma random variables, we directly use the sufficient statistics $(x_a, x_g)$, given in Table 2 of \cite*{damsleth1975conjugate}. The true value of $\alpha$ is 5. The histogram of 5000 posterior samples for each case are shown in Figure \ref{fig:gammashapeunknownscale}. The colored solid lines denote the true posterior density and the black dashed line denotes the true $\alpha$. We see that the posterior samples are indeed sampled from the target distributions. Figure \ref{fig:gamma_trace} shows the sampling trace plots. In Table \ref{tab:gamma_shape_mom}, the sample moments are compared with their theoretical counterparts calculated by numerical integration. When $n=30$, the deviations from the theoretical values are $-1.0\%, -4.6\%, -8.2\%$ and $0.1\%$ for mean, variance, skewness and kurtosis respectively.

\begin{table}[h!]
	\centering
	\caption{Posterior Moments}
	\label{tab:gamma_shape_mom}
	\begin{tabular}{@{}ccc|ccccc@{}}
		\toprule
	$n$ & $x_a$ & $x_g$ & Method	& Mean   & Variance    & Skewness & Kurtosis  \\ \midrule
	\multirow{2}{*}{5}	& \multirow{2}{*}{7.19} & \multirow{2}{*}{6.05} &	Numerical Integration & 4.768 & 5.399 & 0.997   & 4.494 \\
	&	& &	Posterior Sampling & 4.779  & 5.630  & 1.305 &  6.148   \\ \midrule
	\multirow{2}{*}{10}	& \multirow{2}{*}{5.57} & \multirow{2}{*}{5.01} &	Numerical Integration & 6.271 & 5.780 & 0.783   & 3.921 \\
	&	& &	Posterior Sampling & 6.101  & 4.857 & 0.634 &  3.427  \\ \midrule
\multirow{2}{*}{30}	& \multirow{2}{*}{5.09} & \multirow{2}{*}{4.26} & 	Numerical Integration & 3.252 & 0.585 & 0.490   & 3.361 \\
&	& &	Posterior Sampling    & 3.250 & 0.605 & 0.537    & 3.155 \\ \bottomrule
	\end{tabular}
\end{table}

\begin{figure}
		\centering
		\caption{Optimizing Paths of EM Algorithm}
		\label{fig:gamma_EM}
		\includegraphics[width=0.5\linewidth]{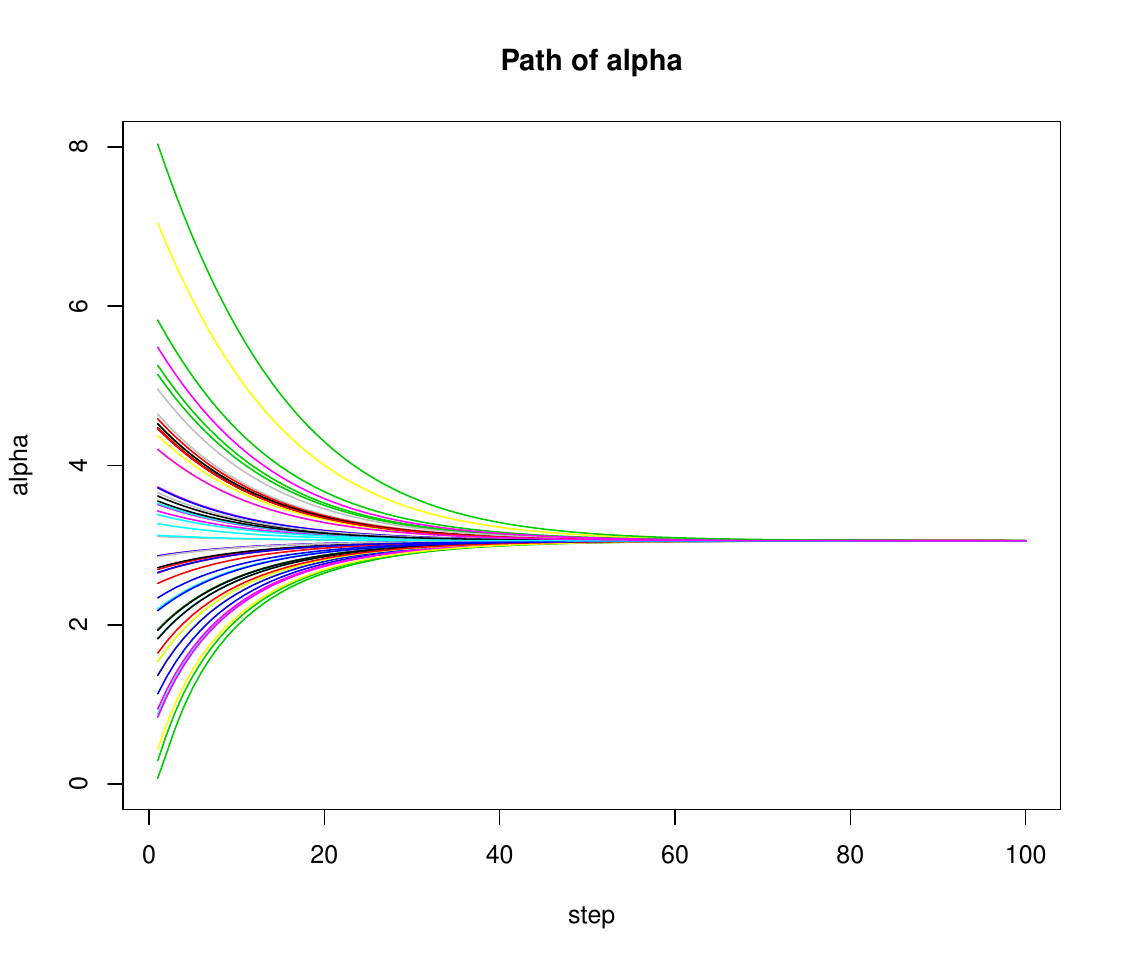}
	\end{figure}

To find the posterior mode of $\alpha$, we exploit the EM algorithm developed in Section \ref{sec:em}. Specially, in the expectation step, the conditional expected value of complete-data log posterior (at $t$-th iteration) given by Equation (\ref{eqn:e-step}) now has the parameters 
	\begin{align*}
	\log Q(x) &:= \delta'\log\alpha\\
	\kappa_1 &:=  \frac{\delta'\left(\gamma + \psi(\alpha^{(t)} + 1)\right)}{2\alpha^{(t)}} > 0\\
	\kappa_2 &:= \delta'\left(\gamma-\log\delta'\eta'/\mu' + \psi(\delta'\alpha^{(t)} + 1)\right)
	\end{align*}
	Then in the maximization step, $\alpha$ is updated by Equation (\ref{eqn:m-step}), with $p-M+1 = \delta'$. In Figure \ref{fig:gamma_EM}, we start with 30 different initial values in EM algorithm and show the optimizing paths for the case $n=30$. In this particular case, the algorithm convP-IGes after around 50 steps. The 30 numerical solutions given by EM has mean 3.054, which matches with the maximizer found by Mathematica. And the standard deviation is as small as 0.0003.   
	
%\subsubsection{Comparison with Deriviative-Matching Method}

\subsection{Negative Binomial Regression}
Next, we proceed to how our data augmentation strategy can be used to fit negative binomial regression models. The count data $\{y_i\}_{i=1}^n$ are assumed to follow the negative binomial distribution
\begin{equation*}
y_i \mid r, p_i \sim \text{NB}(r, p_i),\quad p_i = \frac{1}{1 + e^{-\bm{x_i'\beta}}}
\end{equation*}
where $\{{\bm x_i}\}_{i=1}^n$ are observed covariates and $\bm{\beta}$ is the regression coefficients. $r$ is interpreted as the number of failures until the experiment is stopped, while the success probability $p_i$ is related to $\bm{x_i'\beta}$ via the logistic transformation. This model specification is equivalent to the following Gamma-Poisson mixture,
\begin{align*}
y_i \mid \lambda_i &\sim \text{Poisson}(\lambda_i), \\
\lambda_i \mid r, \bm{x_i'\beta} &\sim \Gamma\left(r, e^{\bm{-x_i'\beta}}\right).
\end{align*}
The likelihood is
\begin{align*}
f(y_i \mid \bm{x_i, \beta}, r) &= \frac{\Gamma(y_i + r)}{y_i! \cdot \Gamma(r)} \left(\frac{e^{\bm{x_i'\beta}}}{e^{\bm{x_i'\beta}} + 1}\right)^{y_i}  \left(\frac{1}{e^{\bm{x_i'\beta}} + 1}\right)^{r}. 
\end{align*}
\cite*{pillow2012fully} adopt exactly the same model with known $r$ and use a data augmentation strategy for the inference on $\beta$. Our example here extends their method and allows for inference on both $r$ and $\beta$. The parameter $r$ controls the dispersion of observations, as the expectation of $y_i$ is $r e^{\bm{x_i'\beta}}$ and variance is $r e^{\bm{x_i'\beta}} \left(1 + e^{\bm{x_i'\beta}}\right)$.

Let the prior for $\bm\beta$ and $r$ be $N(\bm{0}, \Sigma)$ and $p(r)$. We then calculate the joint posterior of $({\bm\beta}, r)$ as 
\begin{align*}
p({\bm \beta}, r \mid {\bm X, y}) &= C\cdot p(r) \exp\left(-\frac{1}{2} \bm{\beta'} \Sigma^{-1} {\bm\beta}\right) \prod_{i=1}^n \frac{\Gamma(y_i + r)}{\Gamma(r)} \left(\frac{e^{\bm{x_i'\beta}}}{e^{\bm{x_i'\beta}} + 1}\right)^{y_i}  \left(\frac{1}{e^{\bm{x_i'\beta}} + 1}\right)^{r} \\
&= C\cdot p(r) \exp\left(-\frac{1}{2} \bm{\beta'} \Sigma^{-1} {\bm\beta}\right) \left(\frac{1}{\Gamma(r)}\right)^n \left(\prod_{i=1}^n  \Gamma(y_i + r) \right)\left(\prod_{i=1}^n \frac{\left(e^{\bm{x_i'\beta}}\right)^{y_i} }{\left(e^{\bm{x_i'\beta}} + 1\right)^{r + y_i}} \right)
\end{align*}
Notice that if we assign gamma prior for $r$, then its conditional posterior density is close to the form (\ref{eqn:post}), except for the extra factor $\prod_{i=1}^n  \left(e^{\bm{x_i'\beta}}\right)^{y_i}/\left(e^{\bm{x_i'\beta}} + 1\right)^{r + y_i}$. Similarly for $\beta$, the posterior is close to normal density, except for the same factor. This is however not an issue as we can write this factor as a scale mixture of normals, where the mixing distribution is P\'olya-Gamma from \cite*{polson2013bayesian}. Setting $a = y_i$ and $b = r+y_i$, the key mixture representation in Equation (\ref{eqn:pg integral}) related to it now becomes
\begin{align*}
\frac{\left(e^{\bm{x_i'\beta}}\right)^{y_i} }{\left(e^{\bm{x_i'\beta} } + 1\right)^{r + y_i}} &\propto \int_0^\infty f(r,\bm{\beta} \mid \xi) \cdot p_{\text{PG}}(\xi_i \mid r+y_i, 0) d\xi_i \\
f(r,\bm{\beta} \mid \xi) &= \exp\left[\frac{1}{2} \left(-r(2\log2 + \bm{x_i'\beta}) + y_i(\bm{x_i'\beta} ) - \xi_i (\bm{x_i'\beta} )^2 \right)\right].
\end{align*}
The integrand $f(r,\bm{\beta} \mid \xi)$ is an exponential density when viewed as a function of $r$, and a normal density when viewed as a function of $\bm\beta$.

Finally, we introduce the auxiliary variables $\bm{\tau, \omega, \xi}$ which follow gamma, P-IG and PG distribution respectively. Let $\Xi := \text{diag}(\xi_1,..., \xi_n)$ and $\bm{z} := \left(\frac{y_1-r}{2\xi_1}, ..., \frac{y_n-r}{2\xi_n}\right)'$ and $p(r) \sim \Gamma(a_0, b_0)$. The conditional posteriors are derived as follows:
\begin{align*}
\tau_i \mid r, \bm{\beta, \xi, \omega, X, y} &\sim \Gamma(y_i + r, 1) \\
\omega_i \mid r, \bm{\beta, \tau, \xi, X, y} &\stackrel{i.i.d.}{\sim}  \text{P-IG}(r) \\
\xi_i \mid r, \bm{\beta, \tau, \omega, X, y} &\sim \text{PG}(r+y_i, \bm{x_i'\beta}) \\
\bm\beta \mid r, \bm{, \tau, \xi, \omega, X, y} &\sim N(\bm m, V) \\
r \mid \bm{\beta, \tau, \xi, \omega, X, y}
&\sim \text{PTN}(a_0+n, a, b+b_0)
\end{align*}
where
\begin{align*}
a &= \sum_{i=1}^n \omega_i\\
b &= n(\gamma-\log2) + \sum_{i=1}^n \left( \log\tau_i - \bm{x_i'\beta}/2 \right)\\
V &= \left(\Sigma^{-1} + X'\Xi X\right)^{-1}\\
m &= VX'\Omega z
\end{align*}

\begin{figure}
	\centering
	\caption{Negative Binomial Regression Results}
	\label{fig:nb}
	\includegraphics[width=0.32\linewidth]{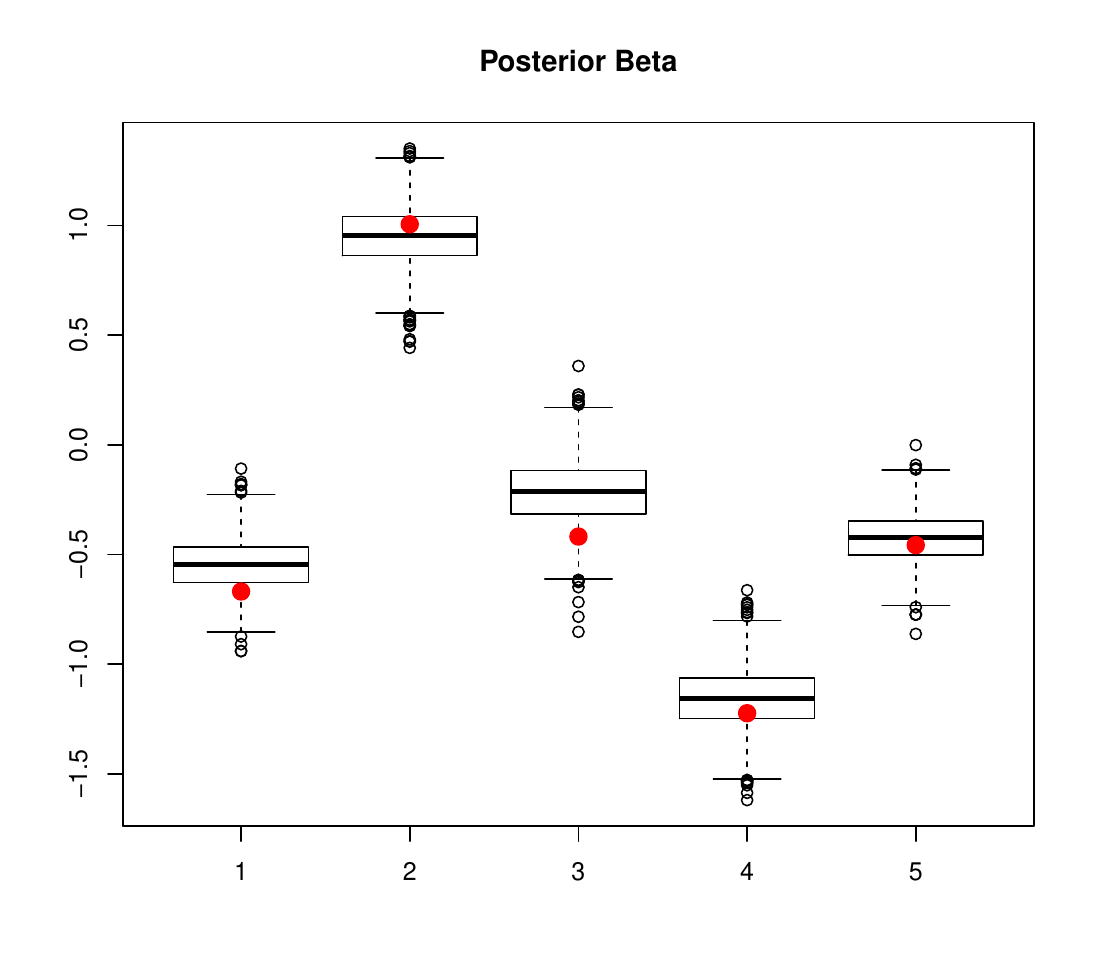}
	\includegraphics[width=0.32\linewidth]{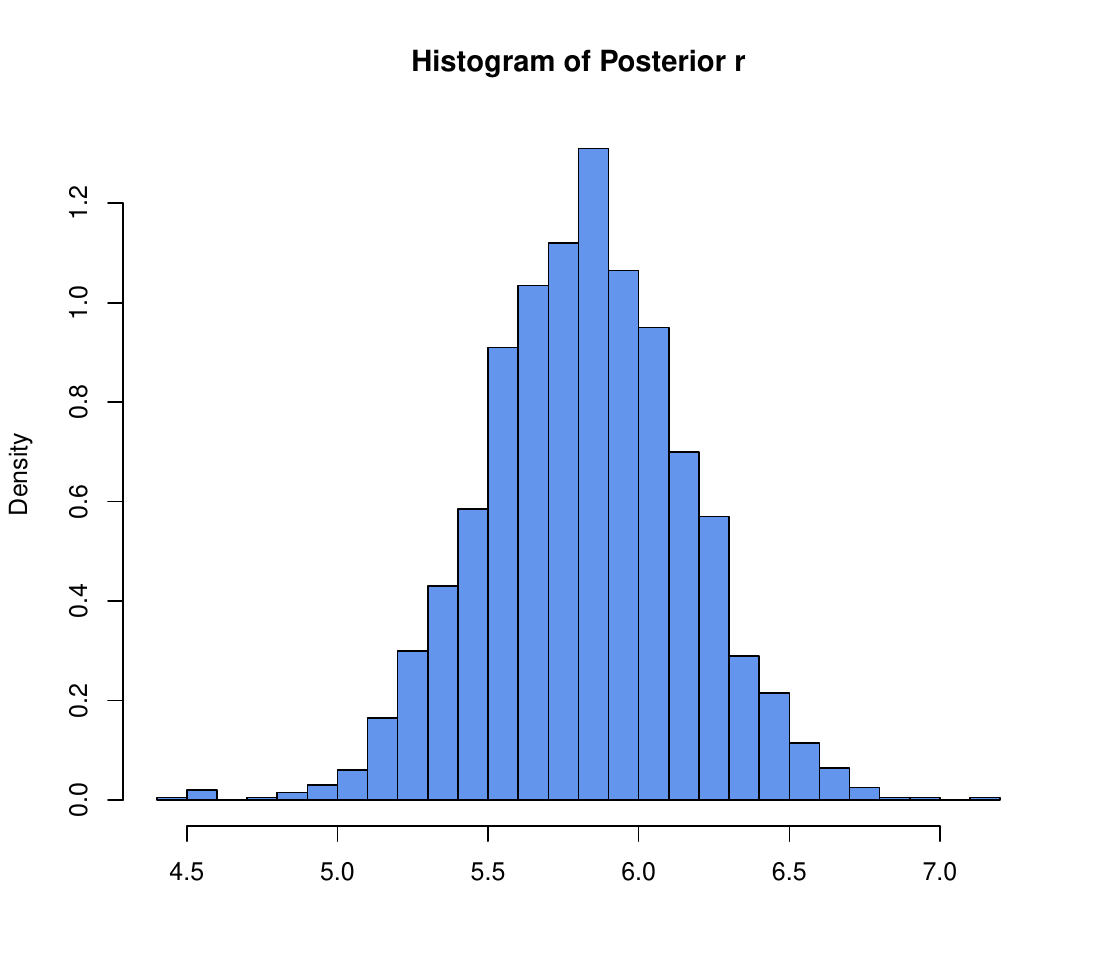}
	\includegraphics[width=0.32\linewidth]{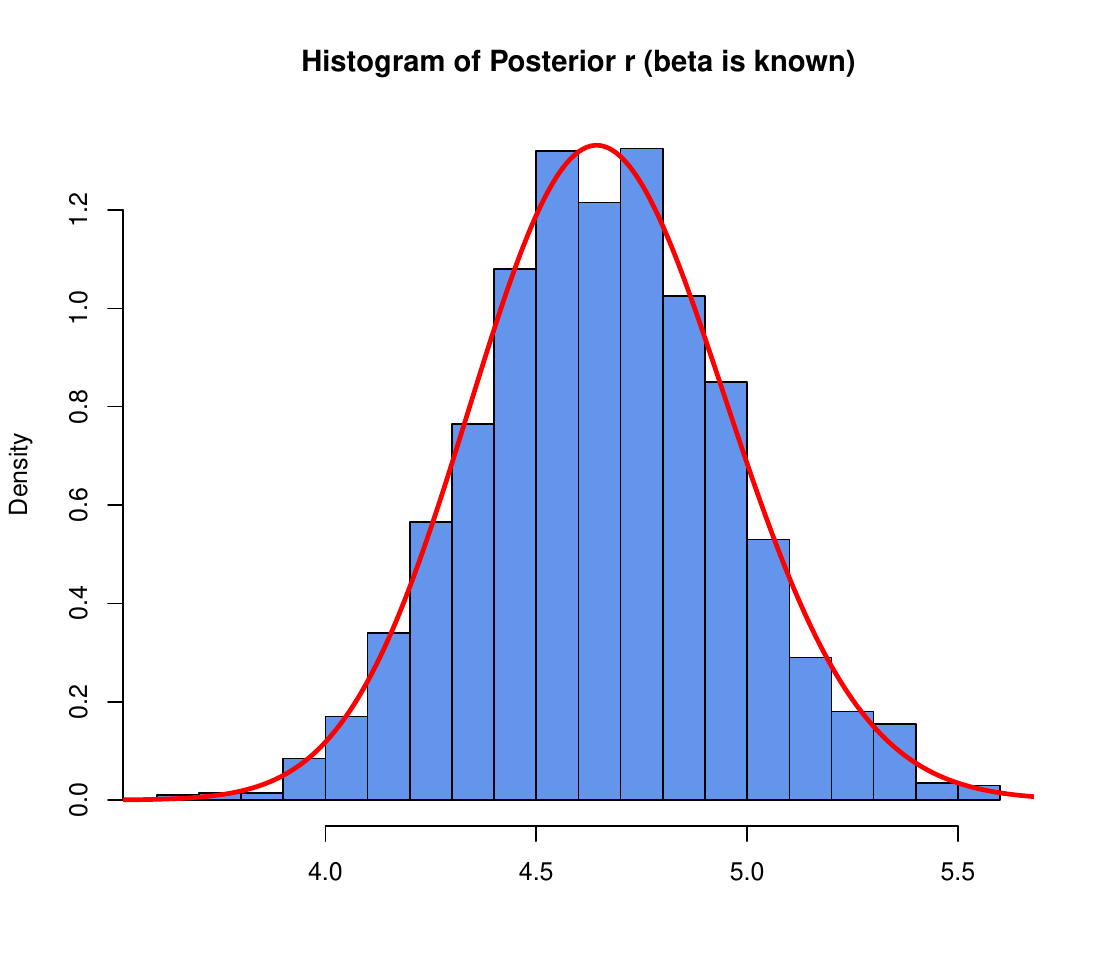}
\end{figure}

Figure \ref{fig:nb} shows an illustrating simulation example where we set true $r = 5$ and generate $n=100$ count observations. We draw the coefficient vector $\bm\beta \in \mathbb{R}^5$ and covariates $x_{ij} \stackrel{i.i.d.}{\sim}  N(0,0.5^2)$ for $1\leq i \leq n$ and $1 \leq j \leq 5$. The prior for $r$ is $p(r)\sim 1/r$ which is the limit case of gamma prior with $a_0=0, b_0=0$. For $\beta$, we choose $\Sigma = 10^6 I_5$ so that the prior information is relatively weak. The left panel shows the boxplots of posterior $\beta$ samples, with those red dots denote the true $\beta$'s. As the figure shows, all 5 $\beta$'s fall in the 95\% credible interval. The middle panel is the histogram of the posterior samples for $r$. Since the actual marginal posterior of $r$ is hard to compute given the complicated form of the joint $p({\bm \beta}, r \mid {\bm X, y})$, in the right panel we use the same dataset, plug in the true $\beta$'s, and run the Gibbs sampler again, so that we are able to compare the resulted histogram with the true density (red line). They are quite close to each other, indicating that the sampling procedure works well for this example. 

\subsection{Multinomial-Dirichlet Model}
In this section, we develop the Markov chain Monte Carlo (MCMC) algorithm for fully posterior inference of the concentration parameter vector in the Dirichlet distribution.  Such inference problems commonly arise in applied analyses of categorical data.  Section \ref{subsec:Model_Framework} presents the general hierarchical multinomial-Dirichlet model class for which the P-IG data augmentation scheme may be utilized.  Section \ref{subsec:Data_Augmentation} develops a Gibbs sampler for inferring the concentration parameter in the Dirichlet distribution and conducts a simulation study comparing the performance of our data augmentation strategy with Metropolis-Hasting algorithm.

\subsubsection{A Hierarchical Multinomial-Dirichlet Model}
\label{subsec:Model_Framework}
The multinomial-Dirichlet framework presented herein is closely related to the latent Dirichlet allocation model of \citet*{blei2003latent} for topic modeling of text data, and we use text analysis as a motivating context.  Suppose that for document $s \in \{1, ..., S\}$, each of $N_s$ words in the document is independently allocated to $K$ topics conditional on probability vector $\bm{p}_s = \left(p_{s1}, p_{s2}, \ldots, p_{sK}\right)$.  For each document $s$, the number of words allocated to each topic is denoted by $\bm{n}_s = \left(n_{s1}, \cdots, n_{sK}\right)$, which follows a multinomial distribution.  The sampling model for the count vector $\bm{n}_s$ is then a multinomial distribution given probability vector $\bm{p}_s$,  
\begin{equation*}
\bm{n}_s \mid N_s, \bm{p}_s  \sim \text{Multinomial}\left(N_s, \bm{p}_s\right).
\end{equation*}
The probability vector $\bm{p}_s$ is the proportional allocation of each document to the $K$ topics.  In a Bayesian analysis, the probability vector for each document $\bm{p}_s$ is typically assigned a Dirichlet distribution with concentration parameter vector $\bm{\alpha} = \left(\alpha_1, \ldots, \alpha_K\right)$, 
\begin{align*}
\bm{p}_s \mid \bm{\alpha} \sim \text{Dirichlet}\left(\bm{\alpha}\right).
\end{align*}
Rather than fixing $\bm{\alpha} = \left(\frac{1}{K}, \ldots, \frac{1}{K}\right)$, as is common, we complete the model with a prior distribution $p(\bm{\alpha})$. This hierarchical prior distribution for $\bm{\alpha}$ facilitates more efficient information sharing across documents (observational units), and it yields practical advantages for out-of-sample prediction, which we discuss below.  The model framework and P-IG augmentation admit independent uniform, truncated normal, and exponential prior distributions for the elements $\alpha_k$.  Section \ref{subsec:Data_Augmentation} presents analyses based on independent gamma priors $p(\bm{\alpha}) = \prod_{k=1}^K p(\alpha_k)$.

In application, model inferences are often summarized by the posterior predictive distribution for the topic proportion vector $\bm{p}^*$ in a new document. Computing the posterior predictive distribution $p\left(\bm{p}^* \mid \bm{n}_1, ..., \bm{n}_S\right) = \int_{\bm{\alpha}} p\left(\bm{p}^* \mid \bm{\alpha}\right) p(\bm{\alpha} \mid \bm{n}_1, ..., \bm{n}_S)d\bm{\alpha}$ requires posterior computation of $p\left(\bm{\alpha} \mid \bm{n}_1, \ldots, \bm{n}_S\right) \propto p\left(\bm{\alpha}\right) \prod_{s=1}^S p\left(\bm{n}_s \mid \bm{\alpha}\right) $; however, when the probability vectors $\bm{p}_s$ are integrated out of the multinomial likelihood, the marginal likelihood $p\left(\bm{n}_s \mid \bm{\alpha}\right)$ includes elements of $\bm{\alpha}$ inside the gamma function,   
\begin{equation*}
\prod_{s=1}^S p\left(\bm{n}_s \mid \bm{\alpha}\right) = \prod_{s=1}^S \left(\frac{\Gamma\left(\sum_{k=1}^K \alpha_k\right)}{\Gamma\left(\sum_{k=1}^K \left(n_{sk} + \alpha_k\right)\right)} \prod_{k=1}^K \frac{\Gamma\left(n_{sk} + \alpha_k\right)}{\Gamma\left(\alpha_k\right)}\right).
\end{equation*}
Because $\bm{\alpha}$ is nested inside the gamma function, computing $p(\bm{\alpha} \mid \bm{n}_1, \ldots, \bm{n}_S)$ is a challenge.  Previous inference strategies relied on approximations, but in Section \ref{subsec:Data_Augmentation} we introduce a new data augmentation scheme for computing the full posterior $p\left(\bm{\alpha} \mid \bm{n}_1, \ldots, \bm{n}_S\right)$.  

\subsubsection{Data Augmentation and Simulation Study}
\label{subsec:Data_Augmentation}
Assume independent gamma prior distributions for each element of vector $\bm\alpha$ so that $p\left(\bm{\alpha}\right) \propto \prod_{k=1}^K \alpha_k^{a_0-1} e^{-b_0\alpha_k}$, with hyperparamter $a_0$ and $b_0$. Note that gamma priors on each $\alpha_k$ give closed-form full conditional distributions in a Gibbs sampler, which is shown below. When $\alpha_k \sim \Gamma(a_0, b_0)$, where $a_0$ denotes the shape parameter and $b_0$ the rate, the expectation $E[\alpha_k] =a_0/b_0$.  We can set $E[\alpha_k] = 1/K$, a standard choice for the Dirichlet concentration parameter, by choosing $a_0 = b_0/K$.  The prior variance then depends on both the dimension of the Dirichlet distribution, $K$, and the rate parameter, $b_0$.  

We now reorganize the multivariate posterior density of $\bm\alpha$ as
\begin{equation}
p(\bm{\alpha} \mid \bm{n}_1, \ldots, \bm{n}_S) = \left(\prod_{k=1}^K f(\alpha_k)\right) \left(\prod_{s=1}^S \frac{\Gamma(\sum_k \alpha_k)}{\Gamma(\sum_k \alpha_k + N_s)}\right)
\end{equation}
where 
\begin{equation}\label{eqn:falpha}
f(\alpha_k) = \left(\prod_{s=1}^S\Gamma( \alpha_k + n_{sk})\right) \left(\prod_{s=1}^S \frac{1}{\Gamma(\alpha_k)}\right) \alpha_k^{a_0-1}e^{-b_0\alpha_k}.
\end{equation}
Note that for each $\alpha_k$, Equation (\ref{eqn:falpha}) is exactly of the form (\ref{eqn:post}). And the extra factor, $\prod_{s=1}^S \frac{\Gamma(\sum_k \alpha_k)}{\Gamma(\sum_k \alpha_k + N_s)}$, can be replaced with a beta integral representation. A multivariate version of Proposition \ref{prop:cond} produces the conditional posteriors as follows
\begin{align*}
\tau_{sk} \mid \bm{\alpha, \omega, \eta} &\sim \Gamma(\alpha_k+n_{sk}), \\
\omega_{sk} \mid \bm{\alpha, \tau, \eta} &\sim \text{P-IG}(\alpha_k), \\
\eta_s \mid \bm{\alpha, \tau, \omega} &\sim \text{Beta}\left(\sum_{k=1}^K \alpha_k, N_s\right),\\
\alpha_k \mid \bm{\tau, \omega, \eta} &\sim \text{PTN}(S + a_0, a_k, b_k)
\end{align*}
where $a_k=\sum_{s=1}^S \omega_{sk}, b_k = S\gamma - b_0 + \sum_{s=1}^S\log(\tau_{sk}\cdot\eta_s)$. 
Conditioned on the introduced auxiliary variables $(\bm{\tau, \omega, \eta})$, all elements of the vector $\bm{\alpha}$ are now mutually independent, which significantly reduces the difficulty of sampling procedure as we can now sample separately from $K$ univariate distributions. 

The total number of auxiliary variables to be sampled at each iterate is $S(1+2K)$, which may greatly slow down the MCMC algorithm for large values of $S$ and $K$. However, we observe that these variables affect the distribution of $\alpha_k$ only through the parameters $(a_k, b_k)$. The summation form of $(a_k, b_k)$ suggests that we may apply central limit theorem and approximate them by normal variables. Therefore, as mentioned in Section \ref{sec:DA}, the Gibbs sampling algorithm can be approximately simplified to
\begin{enumerate}
	\item Initialize $a_k^{(0)}, b_k^{(0)}$ for $1 \leq k \leq K$.
	\item At step $t$, sample $a_k^{(t)}$ and $b_k^{(t)}$ from $N(\mu_{a,k}, \sigma^2_{a,k})$ and $N(\mu_{b,k}, \sigma^2_{b,k})$ respectively, where
	\begin{align*}
	\mu_{a,k} &= \frac{S}{2\alpha_k}\left(\psi(1+\alpha_k) + \gamma\right)\\ 
	\sigma_{a,k}^2 &= \frac{\mu_{a,k}}{2\alpha_k^2} - \frac{S}{4\alpha_k^2}\psi'(1+\alpha_k)\\
	\mu_{b,k} &= S\gamma- b_0+ S\psi(\alpha_0)  + \sum_{s=1}^S \psi(\alpha_k + n_{sk}) - \psi(\alpha_0+N_s) \\
	\sigma_{b,k}^2 &= S\psi'(\alpha_0) + \sum_{s=1}^S \psi'(\alpha_k +n_{sk}) - \psi'(\alpha_0 + N_s)
	\end{align*}
	and $\alpha_0 = \sum_{k=1}^K \alpha_k$.
	\item Sample $\alpha_k^{(t)}$ from $\text{PTN}(S+a_0, a_k^{(t)}, b_k^{(t)})$ for $1 \leq k \leq K$. Increase $t$ by 1 and return to (2).
\end{enumerate}

Fixing the dataset dimensions $S$ and $K$, we consider two settings of the true $\bm\alpha$ for our simulation experiment: (A) heterogeneous $\alpha_k = k/K$ and (B) homogeneous $\alpha_k = 1/K$ for $k=1,2, ..., K$. Probability vector $\bm{p}_s$ are drawn independently from $\text{Dirichlet}(\bm{\alpha})$ and vector of counts $\bm{n}_s$ are drawn from $\text{Mutinomial }(N_s, \bm{p}_s)$ with $N_s=500$ for $s=1,2, ..., S$. We set the hyperparamters $a_0 = b_0/K$ in independent gamma priors of $\alpha_k$'s, so that the prior expectations are all equal to $1/K$. The setting of our simulation experiment is summarized below:
\begin{itemize}
	\item $S\in \{100, 1000\}, K\in \{10, 50\}$ and $b_0\in \{0.1, 1, 5\}$.
	\item 4 MCMC algorithms in comparison:
	\begin{itemize}
		\item DA: Gibbs sampler which iteratively samples $\bm{\tau, \omega, \eta}$ and $\bm\alpha$.
		\item DA-N: replace the parameters in the conditional posterior of $\bm\alpha$ with approximating normal variables and skip the sampling of $\bm{\tau, \omega, \eta}$ in DA.
		\item DA-E: replace the parameters in the conditional posterior of $\bm\alpha$ with corresponding expectations and skip the sampling of $\bm{\tau, \omega, \eta}$ in DA.
		\item MH: random-walk Metropolis-Hasting sampler.
	\end{itemize}
    \item Metrics:
    \begin{itemize}
    	\item Root Mean Square Error: $$\text{RMSE} = \sqrt{\frac{1}{TK}\sum_{t=1}^{T} \Vert \bm{\alpha}^{(t)} - \bm{\alpha} \Vert^2 }.$$ $T=500$ is the posterior sample size. 
    	\item Effective sample size ratio (ESSR): $$\text{ESSR} = \frac{1}{K}\sum_{k=1}^K\frac{\lambda_k^2}{\sigma_k^2},$$ which measures the serial correlation between posterior samples. $\lambda^2$ is the sample variance and $\sigma^2$ is the estimate of the spectral density at frequency zero. 
    	\item Algorithm running time in seconds
    \end{itemize}
\end{itemize}

In simulation experiments, we initialize each $\alpha_k$ with a random draw from $\text{Lognormal}(0,1/K)$ and multiply by $1/K$, so that the prior median is $1/K$. For each combination of $(b_0, S, K)$, we run the simulation for 50 times (the first 200 samples are dropped each time). Table \ref{tab:homo} and \ref{tab:hetro} show the results, averaged over 50 runs, for the homogeneous and heterogeneous setting respectively. Despite being slow, our data augmentation strategy with P-IG auxiliary variables produces more accurate estimates of $\bm\alpha$ than Metropolis-Hasting does, in heterogeneous setting. While in homogeneous setting, the two methods have similar RMSE. Metropolis-Hasting gets significantly worse when $K$ grows to 50. This is not surprising at all. As we notice that, the effective sample size ratios for Metropolis-Hasting are as low as 0.01, indicating strong serial correlations in posterior samples drawn by Metropolis-Hasting, of which the acceptance rate is around 0.02. Therefore, the RMSE for Metropolis-Hasting is entirely up to the distance between the initial $\bm{\alpha}_{(0)}$ and the true $\bm\alpha$. Our Gibbs sampler instead enjoys the advantage of being less sensitive to initialization and not requiring further tuning. The different choices of $b_0$ don't seem to affect the posteriors too much. Once replacing auxiliary variables with normal approximations or expectations, the Gibbs sampler gets much faster. Meanwhile, root mean squared errors and effective sample sizes get slightly improved. With DA-N and DA-E, we can expect the resulted posterior samples to have slightly less variation as well as less correlation.

\begin{table}[h!]
%	\resizebox{0.8\columnwidth}{!}{%
	\centering
	\caption{Homogeneous Setting}
	\small
	\label{tab:homo}
	\begin{tabular}{@{}cc|cccc|cccc|cccc@{}}
		\toprule
		&    & \multicolumn{4}{c}{$b_0=0.1$}    & \multicolumn{4}{c}{$b_0=1$}      & \multicolumn{4}{c}{$b_0=5$}      \\ \midrule
		&    & \multicolumn{12}{c}{Root Mean Square Error ($\times 10^3$)}                                                            \\ \midrule
		$S$    & $K$  & DA    & DA-N  & DA-E  & MH    & DA    & DA-N  & DA-E  & MH    & DA    & DA-N  & DA-E  & MH    \\ \midrule
		100  & 10 & 20.93 & 20.95 & 18.95 & 20.56 & 20.25 & 20.36 & 18.31 & 19.99 & 20.22 & 20.25 & 18.38 & 20.19 \\
		100  & 50 & 7.78  & 8.03  & 6.93  & 5.88  & 7.71  & 7.95  & 6.86  & 6.03  & 7.66  & 7.95  & 6.84  & 5.95  \\
		1000 & 10 & 6.54  & 6.53  & 5.95  & 7.13  & 6.58  & 6.54  & 6.01  & 7.26  & 6.38  & 6.36  & 5.80  & 6.87  \\
		1000 & 50 & 2.49  & 2.51  & 2.19  & 2.48  & 2.49  & 2.51  & 2.20  & 2.50  & 2.48  & 2.50  & 2.18  & 2.46  \\ \midrule
		&    & \multicolumn{12}{c}{Effective Sample Size Ratio}                                                                     \\ \midrule
		$S$    & $K$  & DA    & DA-N  & DA-E  & MH    & DA    & DA-N  & DA-E  & MH    & DA    & DA-N  & DA-E  & MH    \\ \midrule
		100  & 10 & 0.32  & 0.33  & 0.32  & 0.04  & 0.32  & 0.32  & 0.32  & 0.04  & 0.32  & 0.32  & 0.32  & 0.04  \\
		100  & 50 & 0.08  & 0.08  & 0.07  & 0.01  & 0.08  & 0.08  & 0.07  & 0.01  & 0.08  & 0.08  & 0.07  & 0.01  \\
		1000 & 10 & 0.32  & 0.32  & 0.32  & 0.01  & 0.33  & 0.33  & 0.32  & 0.01  & 0.32  & 0.32  & 0.32  & 0.01  \\
		1000 & 50 & 0.08  & 0.08  & 0.07  & 0.01  & 0.08  & 0.07  & 0.07  & 0.01  & 0.08  & 0.08  & 0.07  & 0.01  \\ \midrule
		&    & \multicolumn{12}{c}{Running Time}                                                                     \\ \midrule
		$S$    & $K$  & DA    & DA-N  & DA-E  & MH    & DA    & DA-N  & DA-E  & MH    & DA    & DA-N  & DA-E  & MH    \\ \midrule
		100           & 10 & 32.76   & 0.63  & 0.22  & 0.93  & 31.52   & 0.61  & 0.22  & 0.89  & 33.37   & 0.62  & 0.23  & 0.89  \\
		100           & 50 & 174.29  & 3.61  & 1.12  & 2.51  & 166.49  & 3.43  & 1.08  & 2.43  & 165.67  & 3.43  & 1.07  & 2.45  \\
		1000          & 10 & 236.33  & 5.13  & 1.31  & 9.47  & 206.65  & 4.59  & 1.21  & 8.70  & 204.72  & 4.62  & 1.22  & 8.79  \\
		1000          & 50 & 1099.74 & 30.67 & 6.18  & 26.05 & 1012.35 & 28.12 & 5.77  & 23.99 & 1014.13 & 28.14 & 5.75  & 23.90  \\ \bottomrule
	\end{tabular}
%	}
\end{table}

\begin{table}[h!]
	\centering
	\caption{Heterogeneous Setting}
	\small
	\label{tab:hetro}
	\begin{tabular}{@{}cc|cccc|cccc|cccc@{}}
		\toprule
		&    & \multicolumn{4}{c}{$b_0=0.1$}    & \multicolumn{4}{c}{$b_0=1$}      & \multicolumn{4}{c}{$b_0=5$}      \\ \midrule
		&    & \multicolumn{12}{c}{Root Mean Square Error ($\times 10^3$)}                                                            \\ \midrule
		$S$    & $K$  & DA    & DA-N  & DA-E  & MH    & DA    & DA-N  & DA-E  & MH    & DA    & DA-N  & DA-E  & MH    \\ \midrule
		100           & 10 & 82.76  & 82.25 & 76.54 & 82.34  & 82.56 & 81.90 & 77.33 & 83.49  & 81.95 & 82.11 & 78.46 & 86.96  \\
		100           & 50 & 81.50  & 81.41 & 75.83 & 553.05 & 78.53 & 78.39 & 73.47 & 553.83 & 81.99 & 81.98 & 79.15 & 553.38 \\
		1000          & 10 & 26.11  & 26.00 & 24.50 & 41.00  & 25.76 & 25.61 & 24.18 & 41.08  & 25.88 & 25.71 & 24.25 & 41.39  \\
		1000          & 50 & 24.80  & 24.79 & 23.19 & 553.35 & 25.22 & 25.11 & 23.66 & 553.30 & 25.26 & 25.21 & 23.71 & 552.86 \\ \midrule
		&    & \multicolumn{12}{c}{Effective Sample Size Ratio}                                                                     \\ \midrule
		$S$    & $K$  & DA    & DA-N  & DA-E  & MH    & DA    & DA-N  & DA-E  & MH    & DA    & DA-N  & DA-E  & MH    \\ \midrule
		100           & 10 & 0.47   & 0.49  & 0.50  & 0.03   & 0.48  & 0.51  & 0.49  & 0.03   & 0.49  & 0.51  & 0.51  & 0.03   \\
		100           & 50 & 0.44   & 0.45  & 0.44  & 0.01   & 0.44  & 0.45  & 0.44  & 0.01   & 0.45  & 0.46  & 0.46  & 0.01   \\
		1000          & 10 & 0.48   & 0.50  & 0.50  & 0.02   & 0.47  & 0.50  & 0.50  & 0.01   & 0.47  & 0.50  & 0.50  & 0.01   \\
		1000          & 50 & 0.44   & 0.44  & 0.44  & 0.01   & 0.44  & 0.45  & 0.45  & 0.01   & 0.44  & 0.44  & 0.45  & 0.01   \\ \midrule
		&    & \multicolumn{12}{c}{Running Time}                                                                     \\ \midrule
		$S$    & $K$  & DA    & DA-N  & DA-E  & MH    & DA    & DA-N  & DA-E  & MH    & DA    & DA-N  & DA-E  & MH    \\ \midrule
		100           & 10 & 42.28   & 0.58  & 0.26  & 0.93   & 32.97   & 0.49  & 0.23  & 0.85   & 33.36   & 0.49  & 0.24  & 0.86   \\
		100           & 50 & 186.00  & 3.07  & 1.23  & 2.48   & 164.91  & 2.75  & 1.11  & 2.25   & 164.60  & 2.75  & 1.11  & 2.26   \\
		1000          & 10 & 230.54  & 3.69  & 1.27  & 8.92   & 205.64  & 3.33  & 1.18  & 8.41   & 204.56  & 3.34  & 1.18  & 8.51   \\
		1000          & 50 & 1104.85 & 22.43 & 6.25  & 24.09  & 1014.25 & 21.06 & 5.85  & 22.41  & 1013.74 & 21.06 & 5.86  & 22.46  \\ \bottomrule
	\end{tabular}
\end{table}

\section{Discussion}\label{sec:discussion}
The class of P\'olya Inverse Gamma (P-IG) distributions are developed as mixing distributions for models with Gamma functions. 
This adds to the literature on normal variance-mean mixtures by showing that they extend to a wide class of applications. 
Our ensuing data augmentation strategy facilitates full posterior inference for parameters in models which were hitherto hard to provide inference.  
The algorithms are  scalable and are a fast efficient simulation method for drawing from posterior distributions with applications to many area, such as non-parametric Bayes, latent Dirichlet allocation, Gamma-Gamma hierarchical models, extreme value models, and many other Bayesian mixture models.

The focus of our  paper is on theoretical and algorithmic development of P-IG auxiliary variables.  Our work contributes to the literature on scale mixtures of normals (see, e.g., \citep*{andrews1974scale, west1987scale, polson2013bayesian}).  We believe that the computational strategies developed here will provide the foundation for new and richly structured hierarchical gamma models. Applied Bayesian analyses of categorical data will benefit from increased model flexibility and information borrowing strategies.       

There are a number of avenues for future research. In particular, regularized scale allocation models can be implemented using data augmentation methods of \cite*{polson2013data} with P-IG distribution. \cite*{barndorff1992multivariate} provide multivariate GIG distribution theory and relationships with Poisson processes.

\bibliography{PGgamma} 

\newpage
\appendix
\section{Proof of Theorem \ref{thm:theorem_1}}\label{appendix:proof}
The generalized inverse Gaussian distribution, $\text{GIG}(p, a, b)$,  has  probability density function 
$$
p\left(x\right) \propto x^{p - 1}\exp\left\{-\frac{1}{2}\left(a x + b/x\right)\right\}, \quad a,b,x > 0, p\in\mathbb{R}.$$ 
It suffices to show that if $G_k\sim GIG\left(-\frac{3}{2}, 2c^2, \frac{1}{2k^2}\right)$, then the following integral identity holds, 
$$
E(e^{-s^2 G_k}) = \left( \frac{k + \sqrt{s^2 + c^2}}{k + c} \right) e^{-\frac{\sqrt{s^2 + c^2}-c}{k}}.
$$
The density of $G_k$ given by
$$
p_{k, c}(x) = m\left(k, c\right) x^{-\frac{5}{2}} \exp\left(-\frac{1}{4k^2x}  - c^2x\right).
$$ with normalizing constant, $$m(k, c) = \frac{1}{\Gamma\left(\frac{3}{2}\right)}\frac{ (2k)^{-3}}{c k^{-1} +1}e^{c k^{-1}}.$$ 
It follows by the algebraic calculation,
\begin{align*}
\int_0^\infty e^{-t^2 x} p_{k, c}(x) dx &= m(k, c) \int_0^\infty  x^{-\frac{5}{2}} \exp\left(-\frac{1}{4k^2} x^{-1} -(t^2 + c^2) x\right) dx \\
&= \frac{m(k, c)}{m\left(k, \sqrt{t^2 + c^2}\right)}\\
&= \frac{(\sqrt{t^2 + c^2} k^{-1} +1) \exp\left(\sqrt{t^2 + c^2} k^{-1}\right)}{(c k^{-1} +1)\exp( ck^{-1})}\\
&=  \left( \frac{k + \sqrt{t^2 + c^2}}{k + c} \right) e^{-\frac{\sqrt{t^2 + c^2}-c}{k}},
\end{align*}
as required.

\section{Simulating P-IG Random Variables}\label{appendix:P-IG}
We consider below 3 different ways to generate independent random variables from P-IG distribution.
\begin{enumerate}[label=(\alph*)]
	\item\label{method_1} Since Theorem \ref{thm:theorem_1}, we can approximate an P-IG$(c)$ with the finite sum
	\begin{equation}
	X_N = \sum_{k=1}^{N-1} \text{GIG}\left(-\frac{3}{2}, 2c^2, \frac{1}{2k^2}\right) + \Gamma(a_N,b_N)
	\end{equation}
	where the gamma random variable $\Gamma(a_N,b_N)$ are to approximate the tail part by matching the first two moments given in Theorem \ref{thm:moments}. The shape and rate parameter are 
	\begin{align*}
	b_N &= \frac{2c^2\left(\psi(N+c) - \psi(N)\right)}{\psi(N+c) - \psi(N) - c \psi'(N+c)}\\
	a_N &= \frac{b_N}{2c} \left(\psi(N+c) - \psi(N)\right)
	\end{align*} 
	
	\item Since the generation of i.i.d. inverse gamma variables is faster than that of GIG variables, for small values of $c$, we may consider rejection sampling with P-IG(0) as the proposal density.
	\begin{enumerate}
		\item[1.] Generate a sample $W$ from P-IG(0) using the method in \ref{method_1} and $U$ from Unif[0,1].
		\item[2.] If $U < e^{-c^2W}$, accept $W$ as a sample drawn from P-IG$(c)$. Otherwise, reject $W$ and return to the sampling step. 
	\end{enumerate}
	\item \cite*{mcleish2014simulating} shows that one can simulate random variables using the saddlepoint approximation, given the cumulant generating function $k(t) = \log E(e^{tW})$ is known. The saddlepoint approximation is 
	\begin{align*}
	P(W \leq x) &\approx \Phi(w) + \phi(w)\left(\frac{1}{w} - \frac{1}{u}\right) \\
	w &= w(t) = \text{sgn}(t)\sqrt{2(tk'(t)-k(t))}\\
	u &= u(t) = t\sqrt{k''(t)}  
	\end{align*}
	where $t$ solves $k'(t)=x$. $\Phi(\cdot)$ and $\phi(\cdot)$ are the cdf and density of standard normal distribution. We can first generate a random variable $T$ using inverse transform method, such that its cdf $F(t) = \Phi(w(t)) + \phi(w(t))\left(\frac{1}{w(t)}-\frac{1}{u(t)}\right)$, then $W = k'(T)$ has cdf given by the saddlepoint approximation above. 
	\begin{enumerate}
		\item[1.] Genearte a sample $U$ from Unif[0,1].
		\item[2.] Solve $U = F(t)$ using Newton-Raphson iteration
		\begin{align*}
		t_{n+1} &= t_n - \frac{F(t_n) - U}{\sqrt{k''(t_n)}\phi(w(t_n))}\\
		k(t) &= -\gamma\sqrt{c^2-t} - \ln\Gamma(1+\sqrt{c^2-t}) + \left(\gamma c + \ln\Gamma(1+c)\right)\\
		k'(t) &= \frac{\gamma + \psi(1+\sqrt{c^2-t})}{2\sqrt{c^2-t}}\\
		k''(t) &= \frac{\gamma + \psi(1+\sqrt{c^2-t}) - \sqrt{c^2-t}\psi'(1+\sqrt{c^2-t})}{4(c^2-t)^{3/2}}
		\end{align*}
		\item[3.] $W = k'(t)$.
	\end{enumerate}
	
\end{enumerate}

\section{Simulating PTN Random Variables}\label{appendix:PTN}
The probability density of a random variable $\text{PTN}(p, a, b)$ is given as 
$$
p(x \mid p,a,b) = \frac{x^{p-1}e^{-ax^2+bx}}{\int_0^\infty x^{p-1}e^{-ax^2+bx}dx},\quad  (x, p,a >0, b\neq 0).
$$
Note that we can write it as a multiplication 
\begin{align*}
p(x \mid p,a,b) &=   \frac{x^{p-1}e^{-ax^2+bx}}{\int_0^\infty x^{p-1}e^{-ax^2+bx}dx} \\
&=  \frac{x^{p-1}e^{-(\tau|b|-b)x}e^{-a(x - \tau|b|/2a)^2}}{\int_0^\infty x^{p-1}e^{-(\tau|b|-b)x}e^{-a(x - \tau|b|/2a)^2}dx}\\
& = ce^{-a(x - \tau|b|/2a)^2}g(x \mid p,\tau|b|-b) 
\end{align*}
where $g(x|p,b)$ is the density of a gamma random variable with shape $p$ and rate $\tau|b|-b > 0$. $$0\leq e^{-a(x - \tau|b|/2a)^2} \leq 1,\, c =\frac{\int_0^\infty x^{p-1}e^{-(\tau|b|-b)x}dx}{\int_0^\infty x^{p-1}e^{-(\tau|b|-b)x}e^{-a(x - \tau|b|/2a)^2}dx} \geq 1$$
We sample from $p(x \mid p,a,b)$ by rejection method: 
\begin{enumerate}
	\item Generate $X\sim \Gamma(p, \tau|b|-b)$ and $U\sim Unif[0,1]$
	\item Return $X$ until $U \leq e^{-a(X - \tau|b|/2a)^2}$.
\end{enumerate}
If $b>0$, we set $\tau = \sqrt{1/4 + 2ap/b^2} + 1/2$; if $b<0$, we set $\tau = \sqrt{1/4 + 2ap/b^2} - 1/2$. Then $e^{-a(E(X) - \tau|b|/2a)^2} = 1$.

\section{Approximating Gibbs Sampler} \label{appendix:aGibbs}
If $g_\ell(x)=g(x)$ for all $\ell$ (similarly for $h$ and $j$), and $L,M,N$ are all very large, then an approximate Gibbs sampler based on the conditional posterior in Proposition \ref{prop:cond} is
\begin{enumerate}
	\item Initialize $\tilde{a}^{(0)}, \tilde{b}^{(0)}$.
	\item At step $t$, sample $(\tilde{a}^{(t)}, \tilde{b}^{(t)})$ from $N(\mu_a(t), \sigma^2_a(t))$ and $N(\mu_b(t), \sigma^2_b(t))$ respectively, where
	\begin{align*}
	\mu_a(t) &= a + \frac{M(h'(0))^2}{2h(x^{(t-1)})} \left(\psi(1+h(x^{(t-1)})) + \gamma\right)\\
	\mu_b(t) &= b + Lg'(0)\psi(g(x^{(t-1)})) + M\gamma h'(0) - 2(\mu_a(t)-a) + N j'(0)\left(\psi(j(x^{(t-1)})) - \psi(j(x^{(t-1)})+\beta_n)\right)\\
	\sigma_a^2(t) &= \frac{M(h'(0))^4 }{4(h(x^{(t-1)}))^3} \left(\psi(1+h(x^{(t-1)}))+\gamma-h(x^{(t-1)})\psi'(1+h(x^{(t-1)}))\right)\\
	\sigma_b^2(t) &= L(g'(0))^2\psi'(g(x^{(t-1)})) + 4(h'(0))^2\sigma_a^2(t) + N(j'(0))^2 \left(\psi'(j(x^{(t-1)})) - \psi'(j(x^{(t-1)})+\beta_n)\right)
	\end{align*}
	\item Sample $x^{(t)}$ from $\sum_{m=0}^M \pi_k \cdot \text{PTN}(p+m, \tilde{a}^{(t)}, \tilde{b}^{(t)})$.
\end{enumerate} 
\end{document}